\pdfoutput=1
\documentclass[
	a4paper,
	article,
]{memoir}
\setbinding{0cm}
\setlrmarginsandblock{2.5cm}{2.5cm}{*} 
\setulmarginsandblock{3cm}{3cm}{*} 
\setheadfoot{\onelineskip}{2.5\baselineskip} 
\setheaderspaces{*}{3\baselineskip}{*} 
\checkandfixthelayout
\setSingleSpace{1.2} 
\SingleSpacing

\newcommand{\affiliation}[1]{\gdef\aff{#1}}
\newcommand{\aff}{}

\copypagestyle{mypagestyle}{headings}
\copypagestyle{title}{mypagestyle}
\nouppercaseheads
\makeevenhead{mypagestyle}{}{}{\thepage}
\makeoddhead{mypagestyle}{}{}{\thepage}
\makeevenfoot{mypagestyle}{}{}{}
\makeoddfoot{mypagestyle}{}{}{}
\makeheadrule{mypagestyle}{\textwidth}{\normalrulethickness}

\makeevenhead{title}{}{}{}
\makeoddhead{title}{}{}{}
\makeevenfoot{title}{}{}{\thepage}
\makeoddfoot{title}{}{}{\thepage}

\usepackage[utf8]{inputenc}
\usepackage{gensymb}

\setsecnumdepth{subsubsection}          
\maxsecnumdepth{subsubsection}

\counterwithout{section}{chapter}       

\usepackage{lmodern} 
\usepackage[T1]{fontenc}
\usepackage{import}

\usepackage[english]{babel}

\usepackage{amsmath, amssymb, amsthm}
\usepackage{bm}
\usepackage{bbm}        
\usepackage{mathtools}
\usepackage{upgreek}
\usepackage{mathrsfs}
\usepackage[absolute,overlay]{textpos}

\usepackage{xcolor}

\usepackage{algorithm}
\usepackage{algpseudocode}

\usepackage{wrapfig}

\theoremstyle{plain}

\theoremstyle{definition}

\usepackage{changes}
\usepackage{todonotes}

\nonzeroparskip
\let\cal\mathcal
\newcommand{\latBoldUp}{\bm} 
\newcommand{\greBoldUp}{\bm} 

\newcommand{\vb}{\latBoldUp{b}}

\newcommand{\vf}{\latBoldUp{f}}

\newcommand{\vp}{\latBoldUp{p}}

\newcommand{\vv}{\latBoldUp{v}}

\newcommand{\vx}{\latBoldUp{x}}

\newcommand{\vF}{\latBoldUp{F}}

\newcommand{\vX}{\latBoldUp{X}}

\newcommand{\vch}{\greBoldUp{\chi}}

\newcommand{\cB}{\cal{B}}

\newcommand{\sH}{\mathscr{H}}

\newcommand{\T}{
			^{\mathrm{T}}
}

\newcommand{\bT}{^{\kern-3pt\mathrm{\scriptscriptstyle T}}} 

\newcommand{\mR}{\mathbb{R}}

\usepackage[
    hypertexnames=false, 
    pdfborder={0 0 0}, 
    pdfencoding=auto, 
    psdextra,
    pdfauthor={Prateek Prateek, Giuseppe Capobianco},
    colorlinks=true,    
    citecolor=blue,     
    linkcolor=blue,     
    urlcolor=blue       
]{hyperref}

\usepackage[
    backend=biber,      
    style=numeric-comp,      
    sorting=none        
]{biblatex}
\AtEveryBibitem{%
    \clearfield{urlyear}%
    \clearfield{urlmonth}%
    \clearfield{urlday}%
}

\usepackage{cleveref}   
\crefname{figure}{Figure}{Figures}
\crefname{table}{Table}{Tables}
\crefname{section}{Section}{Sections}
\crefname{subsection}{Section}{Sections}
\crefname{subsubsection}{Section}{Sections}
\crefformat{equation}{(#2#1#3)}
\Crefformat{equation}{(#2#1#3)}
\crefrangeformat{equation}{(#3#1#4)--(#5#2#6)}
\Crefrangeformat{equation}{(#3#1#4)--(#5#2#6)}

\title{{\bfseries \huge A variational framework for bond-based peridynamics with spatially varying horizons and its asynchronous time integration}}

\author{Prateek Prateek$^{1}$\thanks{Corresponding author: \href{mailto:2404.prateek@fau.de}{2404.prateek@fau.de}}, Giuseppe Capobianco$^{1}$, Kai Partmann$^{2}$, \\ Kerstin Weinberg$^{2}$, Michael Ortiz$^{3}$, Sigrid Leyendecker$^{1,4}$}
\affiliation{
$^{1}$Institute of Applied Dynamics, 
Friedrich-Alexander-Universität Erlangen-Nürnberg, Erlangen, Germany\\
$^{2}$Chair of Solid Mechanics, Universität Siegen, Siegen, Germany\\
$^{3}$Division of Engineering and Applied Sciences, California Institute of Technology, Pasadena, California, USA\\
$^{4}$Faculty of Engineering, School of Mechanical, Medical \& Process Engineering, Queensland University of Technology, Brisbane, Australia}

\date{}
\begin{document}
\pagestyle{mypagestyle}

\maketitle

\begin{abstract}
Bond-based peridynamics provides a non-local framework for modelling fracture without requiring spatial derivatives of the displacement field. However, when spatially varying horizons are used together with non-uniform discretisations, the classical single-horizon bond-based peridynamics formulation leads to asymmetric interactions between material points. These asymmetric interactions violate balance laws and can introduce non-physical artefacts such as ghost forces and spurious wave reflections. In this work, we develop a variational formulation for bond-based peridynamics with spatially varying horizons. Starting from the Lagrange-d'Alembert principle, we derive the governing equations of motion and show that the dual-horizon peridynamics formulation emerges naturally from the variation of the internal energy. 

Building on this variational structure, we construct asynchronous variational integrators that allow different time step sizes in different regions of the domain. This is particularly useful for dynamic fracture simulations with local refinement, where small time steps are required only near regions of high resolution or expected crack growth. Numerical examples involving wave propagation, a pre-cracked plate under tension, and the Kalthoff-Winkler impact experiment demonstrate that the proposed framework removes spurious reflections caused by non-uniform horizons, preserves physically consistent fracture patterns, and achieves results comparable to uniformly refined simulations. At the same time, the asynchronous variational integrator reduces the number of internal force evaluations compared to the standard velocity-Verlet method. The proposed approach therefore provides a consistent variational foundation and an efficient time-integration strategy for bond-based peridynamic simulations with spatially varying horizons.
\end{abstract}


\section{Introduction}

Peridynamics (PD) is a particle-based non-local formulation which offers an alternative to classical continuum mechanics (CCM) for modelling and simulating fracture. It has emerged as a powerful framework where the governing equations do not rely on spatial derivatives of the displacement field. Instead, PD replaces these spatial derivatives with an integral over a finite horizon \cite{sillingReformulationElasticityTheory2000,sillingMeshfreeMethodBased2005,sillingPeridynamicStatesConstitutive2007,madenciPeridynamicTheoryIts2014}. This non-local formulation naturally accommodates discontinuities without requiring specialised techniques such as the extended finite element method (XFEM) or phase-field approaches, which are commonly employed in classical frameworks to handle crack initiation and propagation.

Several formulations exist within the peridynamic framework, distinguished by how the interaction forces between material points are computed. These include bond-based peridynamics, which we will refer to as single-horizon bond-based peridynamics formulation (SHBB-PD) \cite{sillingReformulationElasticityTheory2000,sillingMeshfreeMethodBased2005,haStudiesDynamicCrack2010,gerstlePeridynamicModelingConcrete2007}, ordinary state-based peridynamics (OSB-PD) \cite{sillingPeridynamicStatesConstitutive2007,mitchellNonlocalOrdinaryStatebased2011,leSurfaceCorrectionsPeridynamic2018,liuOrdinaryStatebasedPeridynamic2018}, and non-ordinary state-based peridynamics (NOSB-PD) \cite{warrenNonordinaryStatebasedPeridynamic2009,sillingStabilityPeridynamicCorrespondence2017,chenPeridynamicBondassociatedCorrespondence2019,javaheriHigherOrderApproximationsStabilizing2022}. Among these, bond-based peridynamics remains one of the most widely used models due to its conceptual simplicity, clear physical interpretation, and its ability to model complex fracture phenomena. Despite its known limitations, most notably the restriction on Poisson's ratio \cite{trageserBondBasedPeridynamicsTale2020}, SHBB-PD continues to serve as a fundamental and widely adopted model for fracture simulations.

Classically, single-horizon bond-based peridynamics is formulated through pairwise force interactions and balance of linear momentum, rather than from an explicit variational principle \cite{sillingMeshfreeMethodBased2005,sillingReformulationElasticityTheory2000,sillingPeridynamicStatesConstitutive2007}. In many areas of mechanics, however, governing equations are naturally derived from variational principles such as Hamilton's principle or the principle of virtual work, which provide a systematic and physically consistent foundation. Variational formulations are particularly attractive because they offer a natural pathway towards structure-preserving spatial and temporal discretisations, including geometric integrators with desirable conservation properties. Although microelastic bond-based peridynamic models admit an energetic interpretation, a clear and unified discrete variational framework particularly for formulations with spatially varying horizons has not been systematically developed yet. Establishing such a framework is therefore important both for theoretical understanding and for the construction of geometric numerical methods for peridynamics.

A fundamental aspect of peridynamics is the choice of horizon, which defines the range of non-local interactions between material points. The governing equations of PD are typically derived assuming a uniform horizon, which simplifies both the mathematical formulation and numerical implementation \cite{sillingMeshfreeMethodBased2005,sillingPeridynamicStatesConstitutive2007}. However, in many applications, such as fracture modelling, only a small region requires high resolution, such as near crack tips or near crack surfaces. In such cases, a finer discretisation is desirable locally, whereas a coarser discretisation is sufficient elsewhere. Since the horizon is commonly chosen as a fixed multiple of the local point spacing, non-uniform discretisations naturally lead to spatially varying horizons. This avoids the excessive computational cost of using a globally fine discretisation, and its associated small horizon, throughout the entire domain.

\begin{figure}[!htbp]
    \centering
    \def\svgwidth{\textwidth}
    
    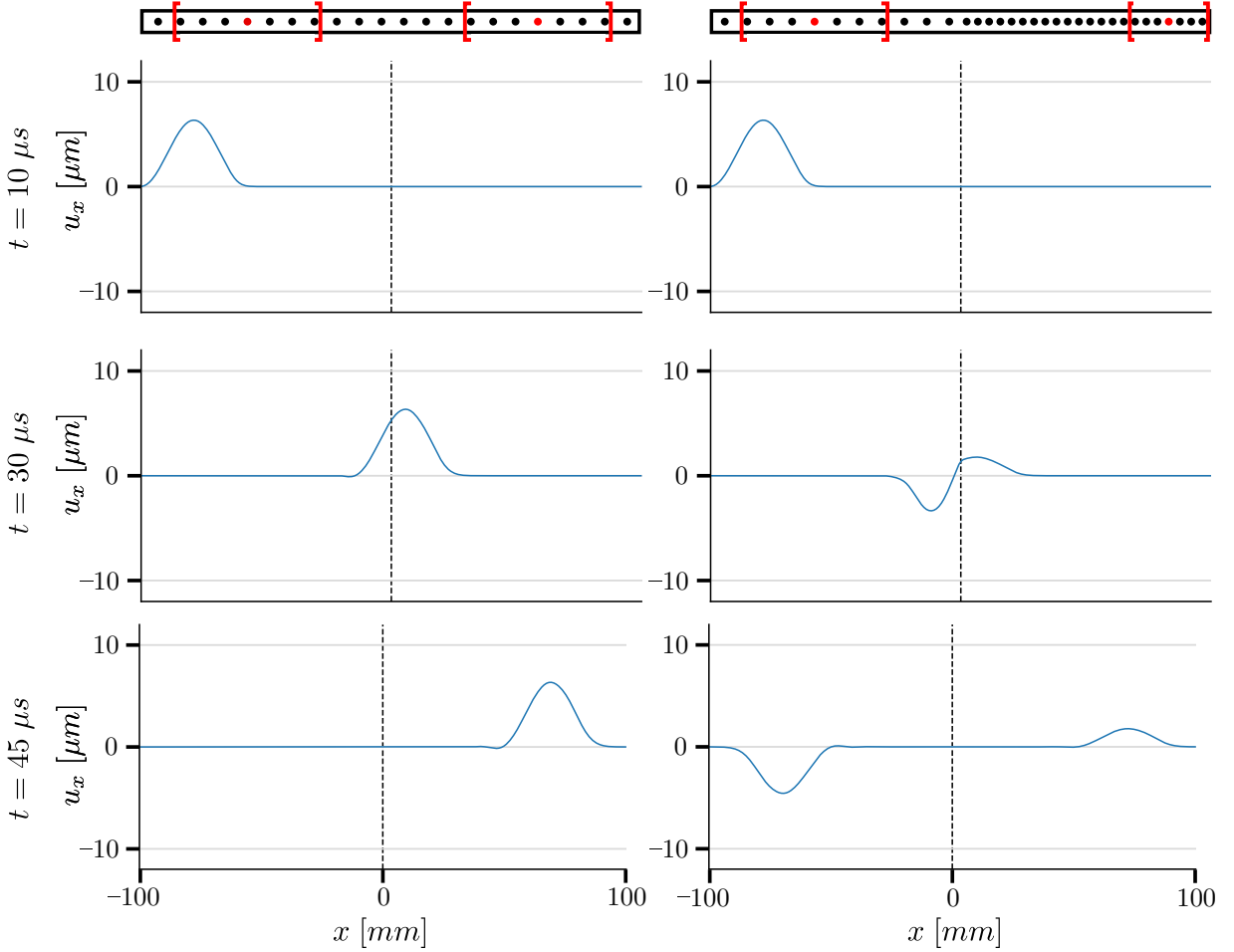

    \caption{Comparison of the effect of discretisation/horizon on wave propagation simulated with SHBB-PD. Left: Bar with a uniform discretisation and uniform horizon. Right: Bar with a non-uniform discretisation and spatially varying horizon in the left and right half sections. A stress wave propagating from the coarse to the fine region generates spurious, non-physical reflections due to asymmetric nonlocal interactions, even in the absence of any material discontinuity, whereas the uniform case does not exhibit any reflections. See \cref{sec:wave_prop_in_a_bar} for details on the problem setup.}
    \label{fig:spurious_reflections}
\end{figure}

The introduction of a non-uniform horizon may lead to several numerical and physical inconsistencies. In particular, the asymmetric interactions between material points violate balance laws and produce non-physical artefacts such as ghost forces and spurious wave reflections \cite{sillingVariableHorizonPeridynamic2015,renDualhorizonPeridynamics2016,renDualhorizonPeridynamicsStable2017}. These reflections arise when stress waves encounter regions with different horizon sizes, which effectively introduce artificial impedance mismatch. As a consequence, part of the wave is reflected at the interface between regions with different horizons, even in the absence of any physical material discontinuity, as shown in \cref{fig:spurious_reflections}. This behaviour deteriorates the accuracy of dynamic fracture simulations and can significantly influence crack propagation predictions. To alleviate these issues, dual-horizon bond-based peridynamics (DHBB-PD) was proposed \cite{renDualhorizonPeridynamics2016,renDualhorizonPeridynamicsStable2017}. In this formulation, each material point is associated with two horizons that ensure symmetric and consistent interactions between neighbouring points, thereby eliminating ghost forces and mitigating spurious reflections. 

Beyond spatial discretisation, time integration is another critical aspect of dynamic fracture simulations. Explicit time integration schemes are commonly used in PD simulations due to their simplicity, ease of implementation and computational efficiency. However, explicit schemes are mostly conditionally stable, and the critical time step is often determined by the smallest horizon in the simulation \cite{littlewoodEstimationCriticalTime2014}. This can lead to prohibitively small time steps when a locally refined discretisation is used. Implicit time integration schemes can alleviate this issue by allowing larger time steps, but they require solving nonlinear systems of equations at each time step, which can be computationally expensive. The possibility of using different time step sizes in different regions of the domain, such as smaller time steps near crack tips and larger time steps away from them, is therefore an attractive approach to reduce computational cost while maintaining accuracy.

Asynchronous Variational Integrators (AVIs) constitute a class of time integration schemes constructed by discretising Hamilton's principle of stationary action in a way that allows for different time step sizes in different regions of the domain \cite{lewAsynchronousVariationalIntegrators2003,lewThesisCaltech2003}. AVIs have successfully been applied to a wide range of problems in mechanics, including solid mechanics \cite{lewThesisCaltech2003,wolffAsynchronousVariationalIntegration2013,leitzVariationalLieGroup2014}, collisions \cite{wolffThesisAVI}, contact mechanics \cite{vougaAsynchronousVariationalContact2011,niuAsynchronousVariationalIntegrator2024}, and phase-field fracture \cite{jadhavNewApproachAsynchronous2025,jadhavSpatiotemporallyAdaptiveAsynchronous2026}. Variational multirate integrators follow a related structure-preserving philosophy by splitting the system into slow and fast variables, which are discretised on coupled macro and micro time grids. Such schemes have been developed for constrained mechanical systems, multibody dynamics, and optimal control \cite{leyendeckerVariationalApproachMultirate2013,gailVariationalMultirateIntegration2016,lishkovaMultirateVariationalApproach2020,ober-blobaumVariationalMultirateIntegrators2024}.

In this work, we develop a unified variational framework for bond-based peridynamics with spatially varying horizons. Starting from a discrete system of interacting particles, we derive the equations of motion using the \textit{Lagrange-d'Alembert principle}. Based on this variational formulation, we construct structure-preserving asynchronous variational integrators that allow for different time step sizes across the domain. Although the derivations are carried out in the context of bond-based peridynamics, the proposed framework can be naturally extended to other peridynamic formulations such as ordinary and non-ordinary state-based models.

The remainder of the paper is organised as follows. \Cref{sec:bond_based_pd_with_variable_horizons} introduces bond-based peridynamics and presents its extension to spatially varying horizons. The micro-elastic constitutive model together with the fracture formulation used throughout the numerical studies is described in \cref{sec:constitutive_model}. In \cref{sec:variational_integrators_for_bbpd}, we develop asynchronous variational integrators grounded in the derived variational formulation of the peridynamic system. Numerical examples are presented in \cref{sec:numerical_examples} to illustrate the effectiveness of the proposed framework in reducing spurious wave reflections and in accurately simulating dynamic fracture phenomena under asynchronous time integration. Finally, conclusions and future research directions are discussed in \cref{sec:conclusions}.

\section{Bond-based peridynamics with Variable Horizons} \label{sec:bond_based_pd_with_variable_horizons}

\begin{figure}[!htbp]           
    \centering                 
    \def\svgwidth{0.67\textwidth}
    
    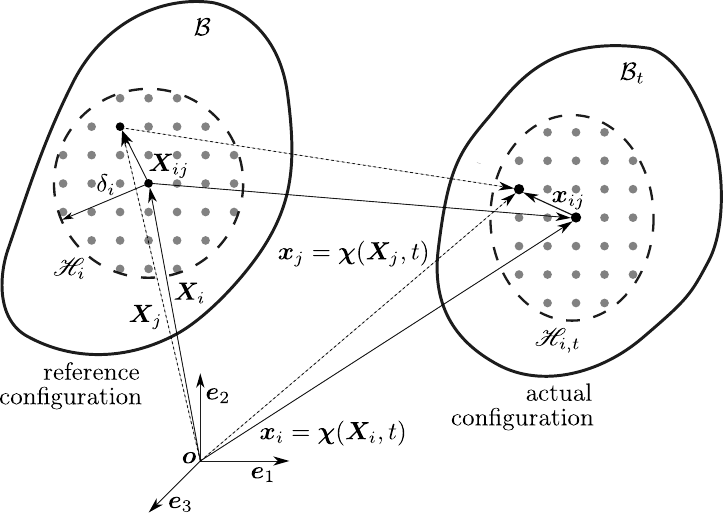
    
    \caption{Representation of discrete peridynamic body.}  
    \label{fig:pd_body}        
\end{figure}

Consider a continuum body occupying a reference domain $\Omega \in \mR^3$. We discretise the domain into $N$ material points forming the set $\cB=\{ \vX_i \in \mR^3 \,|\, i=1, \dots, N\}$ as shown in \cref{fig:pd_body}. The motion of the body is described through the deformation map $\vch\!:\! \cB \times \mR \to \mR^3$, which assigns to each point its spatial position $\vx_i(t)=\vch(\vX_i, t)$ at time $t$. To describe the motion of all the points of the body holistically, we introduce the vector $\vx=(\vx_1, \dots, \vx_N) \in \mR^{3N}$. 

In single-horizon bond-based peridynamics, each point $\vX_i\in\cB$ interacts with all neighbouring points within a spherical region defined by the horizon $\delta_i =  \delta(\vX_i) >0$. The family of points that interact with $\vX_i$ is the index set defined as

\begin{align}
    \sH_{i} = \left\{ j \; \middle| \;  \vX_{j} \in \mathcal{B}, \|  \vX_{j} - \vX_{i}\| \leq \delta(\vX_i)  \right\} .
\end{align}

The actual configuration of the family is denoted by $\sH_{i, t}=\vch(\sH_{i}, t).$ The bond between points $\vX_i$ and each $\vX_j$ in the reference and actual configuration is defined as

\begin{equation}
    \vX_{ij}=\vX_j-\vX_i, \hspace{3cm} \vx_{ij}=\vx_j-\vx_i.
\end{equation}

We derive the equation of motion from the \textit{Lagrange-d'Alembert} principle for dynamical systems. For any curve $\vx: \mR\supseteq [T_0, T_1]\to \mR^{3N}$ the \textit{action} functional $\mathcal{S}$ and \textit{Lagrangian} are defined as

\begin{equation}
    \mathcal{S} [\vx] = \int_{T_0}^{T_1} L(\vx, \dot{\vx}, t) \mathrm{d}t, \hspace{1.5cm} L(\vx, \dot{\vx}, t) = T(\dot{\vx}) - \Pi(\vx,t) ,
\end{equation}

where $T$ is the kinetic energy, and $\Pi$ is the potential energy. The Lagrange-d'Alembert principle states that the actual trajectory of the body satisfies

\begin{equation}\label{eq:principle_of_virtual_work}
    \delta \mathcal{S} [\vx] + \int_{T_0}^{T_1} \delta \vx\T\vf(\vx, t) \, \mathrm{d}t = 0 \quad \forall \, \delta \vx \ \text{with}\ \delta \vx(T_0) = \delta \vx(T_1) = 0.
\end{equation}

Since the peridynamic body is used to approximate a continuum, the material body is subdivided into small voxels of volume $V_i$, which are identified with the centre of mass of the voxels. The mass of each voxel is $m_i = \varrho_i V_i$, where $\varrho_i = \varrho(\vX_i)$ is the mass density. Thus, the kinetic energy of the discretised body is

\begin{equation}\label{eq:kinetic_energy_equation}
    T(\dot{\vx}) = \frac{1}{2} \sum_{i  =1}^{N} m_i \|\dot{\vx}_i\|^2 = \frac{1}{2} \sum_{i  =1}^{N} \varrho_i V_i \|\dot{\vx}_i\|^2\,.
\end{equation}

The total potential energy $\Pi$ consists of the internal $\Pi^{\textrm{int}}$ and external energy $\Pi^{\textrm{ext}}$ contributions, i.e.,

\begin{equation}
    \Pi(\vx, t) = \Pi^{\textrm{int}}(\vx) + \Pi^{\textrm{ext}}(\vx, t)\,.
\end{equation}

In bond-based peridynamics, the internal energy arises from the pairwise interactions between material points. We introduce the \textit{micro-potential} $\pi^{\textrm{int}}_{ij} = \pi^{\textrm{int}}_{ij} \left( \vx_{ij}, \vX_{ij} \right)$ to characterize the energy density of the bond connecting the points $\vX_i$ and $\vX_j$. For bond-based peridynamics, this micro-potential depends only on the relative deformation of the bond, and is independent of the deformation of other bonds.

The total potential energy density contribution of $\sH_i$ is

\begin{equation} \label{eq:pi_i_equations}
    \pi^{\mathrm{int}}_i \left( \vx \right) = \sum_{j \in \sH_i}  \pi^{\mathrm{int}}_{ij} \left( \vx_{ij}, \vX_{ij} \right) V_j.
\end{equation}

With this, we formulate the total internal potential energy within the body as

\begin{equation} \label{eq:internal_energy_expression}
    \Pi^{\mathrm{int}} \left( \vx \right) 
            = \sum_{i=1}^{N} \pi^{\mathrm{int}}_i \left( \vx \right) V_i
            = \sum_{i=1}^{N} \sum_{j \in \sH_i}  \pi^{\mathrm{int}}_{ij} \left( \vx_{ij}, \vX_{ij} \right) V_j V_i .
\end{equation}

External loading stems either from potential forces, which can be derived from the external potential energy $\Pi^{\mathrm{ext}}(\vx, t)$, or from non-conservative forces which are represented by the body force density $\vb^{\mathrm{ext}}_i = \vb^{\mathrm{ext}}(\vx_i, t)$, using which we can write the virtual work done as,

\begin{equation}
    \delta \vx\T\vf = \sum_{i=1}^{N} \delta\vx_i\T \vb_i^{\mathrm{ext}} V_i.
\end{equation}

\subsection{Equation of motion and dual-horizon equivalence} \label{subsec:eqn_motion_and_DH_equivalence}

The variation of the kinetic energy in \cref{eq:kinetic_energy_equation}, after integration by parts and using the endpoint conditions \(\delta \vx(T_0)=\delta \vx(T_1)=0\), becomes

\begin{equation}
    \delta \int_{T_0}^{T_1} T(\dot{\vx}) \, \mathrm{d}t
    =
    - \int_{T_0}^{T_1} \sum_{i=1}^{N} \delta \vx_i\T \, \varrho_i V_i \, \ddot{\vx}_i \, \mathrm{d}t.
\end{equation}

Before computing the variation of $\Pi^{\mathrm{int}}$ in \cref{eq:internal_energy_expression}, we observe that a given index \(i\) appears in the double sum in two distinct roles: first, as the primary index in all terms with \(j \in \sH_i\), and second, as a family member of another point \(j\), that is, whenever \(i \in \sH_j\). To account for the latter contribution systematically, we utilize the dual-horizon set introduced by Ren et al. \cite{renDualhorizonPeridynamics2016,renDualhorizonPeridynamicsStable2017}, which is defined as

\begin{equation}
    \sH'_i = \{ j \; | \; i \in \sH_j \},
\end{equation}

which collects all points with point \(\vX_i\) as one of their neighbour. Using this notation, the variation of the internal potential energy can be written as
\begin{equation}
    \delta \Pi^{\textrm{int}}(\vx)
    =
    \sum_{i=1}^{N}
    \delta \vx_i\T
    \left[
        - \sum_{j \in \sH_i}
        \frac{\partial \pi^{\textrm{int}}_{ij}}{\partial \vx_{ij}} V_j
        +
        \sum_{j \in \sH'_i}
        \frac{\partial \pi^{\textrm{int}}_{ji}}{\partial \vx_{ji}} V_j
    \right] V_i.
\end{equation}

The first summation inside the brackets represents the contribution from all bonds which point \(i\) makes with its neighbours, while the second sum accounts for all bonds for which point \(i\) appears as the neighbour of another point outside the neighbourhood of $i$. Substituting all contributions into \cref{eq:principle_of_virtual_work} yields the Lagrange-d'Alembert principle in the form

\begin{equation}
\begin{aligned}
    \int_{T_0}^{T_1}
    \sum_{i=1}^{N}
    \delta \vx_i\T
    \Bigg[
        - \varrho_i V_i \ddot{\vx}_i
        - \left(
            - \sum_{j \in \sH_i}
            \frac{\partial \pi^{\textrm{int}}_{ij}}{\partial \vx_{ij}} V_j
            +
            \sum_{j \in \sH'_i}
            \frac{\partial \pi^{\textrm{int}}_{ji}}{\partial \vx_{ji}} V_j
        \right) V_i
        - \frac{\partial \Pi^{\textrm{ext}}}{\partial \vx_i}
        + \vb_i^{\textrm{ext}} V_i
    \Bigg]
    \, \mathrm{d}t = 0,
\end{aligned}
\end{equation}

This relation holds for all admissible variations $\delta \vx_i$. Since the variations are otherwise arbitrary, the fundamental lemma of calculus of variations implies that the bracketed term must vanish for every point $i$. Therefore, the equations of motion can be written as

\begin{equation}
    \varrho_i \ddot{\vx}_i
    =
    \sum_{j \in \sH_i}
    \frac{\partial \pi^{\textrm{int}}_{ij}}{\partial \vx_{ij}} V_j
    -
    \sum_{j \in \sH'_i}
    \frac{\partial \pi^{\textrm{int}}_{ji}}{\partial \vx_{ji}} V_j
    -
    \frac{1}{V_i}
    \frac{\partial \Pi^{\textrm{ext}}}{\partial \vx_i}
    +
    \vb_i^{\textrm{ext}}.
\end{equation}

Defining the pairwise internal force density 

\begin{equation} \label{eq:pairwise_force_function_general}
    \vf_{ij} \left( \vx_{ij}, \vX_{ij} \right)
    =
    - \frac{\partial \pi^{\textrm{int}}_{ij}}{\partial \vx_i}
    \left( \vx_{ij}, \vX_{ij} \right)
    =
    \frac{\partial \pi^{\textrm{int}}_{ij}}{\partial \vx_{ij}}
    \left( \vx_{ij}, \vX_{ij} \right),
\end{equation}

and the potential external force density as

\begin{equation}
    \vb_i^{\textrm{ext,c}} = - \frac{1}{V_i} \frac{\partial \Pi^{\textrm{ext}}}{\partial \vx_i}(\vx,t) ,
\end{equation}

where the superscript \textit{c} denotes the contribution from potential forces. The governing equations can be written compactly as

\begin{equation} \label{eq:equation_of_motion_force_form_dual_horizon}
\begin{aligned}
    \varrho_i \ddot{\vx}_i
    =
    \sum_{j \in \sH_i}
    \vf_{ij}\left( \vx_{ij}, \vX_{ij} \right) V_j
    -
    \sum_{j \in \sH'_i}
    \vf_{ji}\left( \vx_{ji}, \vX_{ji} \right) V_j
    +
    \vb_i^{\textrm{ext,c}}
    +
    \vb_i^{\textrm{ext}}.
\end{aligned}
\end{equation}

\textbf{Remark 1 (Dual-horizon equivalence)} The equations derived above coincide exactly with the dual-horizon peridynamics formulation presented by Ren et al. \cite{renDualhorizonPeridynamics2016,renDualhorizonPeridynamicsStable2017}. The DHBB-PD, which was originally introduced to restore momentum balance under spatially varying horizons, here emerges as a natural consequence of the variation of internal potential energy. This provides a rigorous foundation for the dual-horizon approach.

\textbf{Remark 2 (Uniform-horizon case)} When the horizon is constant, $\delta (\vX_i) = \bar{\delta}$ for all $i$, the neighbourhood relationship becomes symmetric, i.e., a constant horizon $\bar{\delta}$ implies

\begin{equation}
    j \in \sH_i \Longleftrightarrow i \in \sH_j .
\end{equation}

Consequently, $\sH_i = \sH'_i$, and hence the two sums in the equation of motion in \cref{eq:equation_of_motion_force_form_dual_horizon} can be combined as

\begin{equation}
\begin{aligned}
        \varrho_i \ddot{\vx}_i = 
        \sum_{j \in \sH_i} \left(
        \vf_{ij}\left( \vx_{ij}, \vX_{ij} \right)
        -
        \vf_{ji}\left( \vx_{ji}, \vX_{ji} \right) \right) V_j + \vb_i^{\textrm{ext,c}} + \vb_i^{\textrm{ext}}.
\end{aligned}
\end{equation}

Since the micro-potential of a bond depends only on the deformation of the material points, the pairwise force function is antisymmetric, i.e. $\vf_{ij} = - \vf_{ji}$. Thus, the equation of motion reduces to

\begin{equation} \label{eq:equation_of_motion_classical_bbpd}
\begin{aligned}
        \varrho_i \ddot{\vx}_i = 
        \sum_{j \in \sH_i} 2 \vf_{ij}\left( \vx_{ij}, \vX_{ij} \right) V_j + \vb_i^{\textrm{ext,c}} + \vb_i^{\textrm{ext}}.
\end{aligned}
\end{equation}

This expression recovers the single-horizon bond-based peridynamic equation of motion, where the extra factor of 2 is typically absorbed into the definition of the bond micro-modulus, and therefore does not appear explicitly in the governing equations \cite{sillingMeshfreeMethodBased2005}.

Next, we illustrate the implications of the DHBB-PD formulation for wave propagation in a bar with non-uniform discretisation and spatially varying horion, and demonstrate how the dual-horizon formulation effectively mitigates spurious wave reflections arising from asymmetric interactions.

\subsection{Wave propagation in a bar} \label{sec:wave_prop_in_a_bar}

A numerical setup similar to the one considered by Partmann et al. \cite{partmannPeridynamicComputationsWave2024} is employed to illustrate the occurrence of spurious wave reflections arising from the use of non-uniform discretisations together with spatially varying horizons in single-horizon bond-based peridynamics, and to demonstrate how the dual-horizon peridynamics formulation effectively mitigates these artefacts. The setup consists of a bar of dimensions $200\,\mathrm{mm} \times 2\,\mathrm{mm} \times 2\,\mathrm{mm}$, which is partitioned into two equal halves along its length. The left and right halves are discretised independently with $N_{x1} \times N_{yz1} \times N_{yz1}$ and $N_{x2} \times N_{yz2} \times N_{yz2}$ material points, respectively. Uniform spacings $\Delta xyz_1$ and $\Delta xyz_2$ are used in the $x$-, $y$- and $z$-directions in each half-section, respectively. The peridynamic horizon in each region is chosen to be proportional to the local discretisation size, such that $\delta_1 = 3.015 \Delta xyz_1$ and $\delta_2 = 3.015 \Delta xyz_2$. 

\begin{figure}[!htbp]
    \centering
    \def\svgwidth{\textwidth}
    
    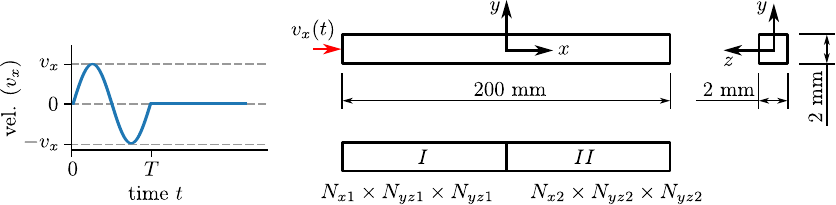

    \caption{Bar setup for wave propagation study with velocity boundary condition on the left face.}

\end{figure}

The material properties of the bar are chosen as $\varrho = 7800\,\mathrm{kg/m^3}$, $E = 210\,\mathrm{GPa}$. A 3D micro-modulus function is used for the simulations, which restricts the Poisson's ratio to $\nu = \frac14$. To generate the stress wave, a time-dependent velocity boundary condition is prescribed on the left face of the bar, inducing a wave propagation along the axial direction. The imposed velocity boundary condition is defined as

\begin{equation}
    v_x(t) = 
    \begin{cases}
        \hat{v}_x \sin \left(\frac{2 \pi}{T} t\right) & \text { if } t \leq T, \\ 0 & \text { else, }
    \end{cases}
\end{equation}

with the velocity amplitude $\hat{v}_x = 2 \,\mathrm{m/s}$ and the period $T = 10\,\mathrm{\mu s}$. To assess the effect of spatially varying discretisations and horizon definitions, three simulation scenarios are considered:

\begin{itemize}
    \item Case (A): uniform discretisation with $N_{yz1} = N_{yz2} = 4$, simulated using SHBB-PD,
    \item Case (B): non-uniform discretisation with $N_{yz1} = 4$ and $N_{yz2} = 5$, simulated using the SHBB-PD, and
    \item Case (C): non-uniform discretisation with $N_{yz1} = 4$ and $N_{yz2} = 5$, simulated using DHBB-PD.
\end{itemize}

\begin{figure}[!htbp]
    \centering
    \def\svgwidth{\textwidth}
    
    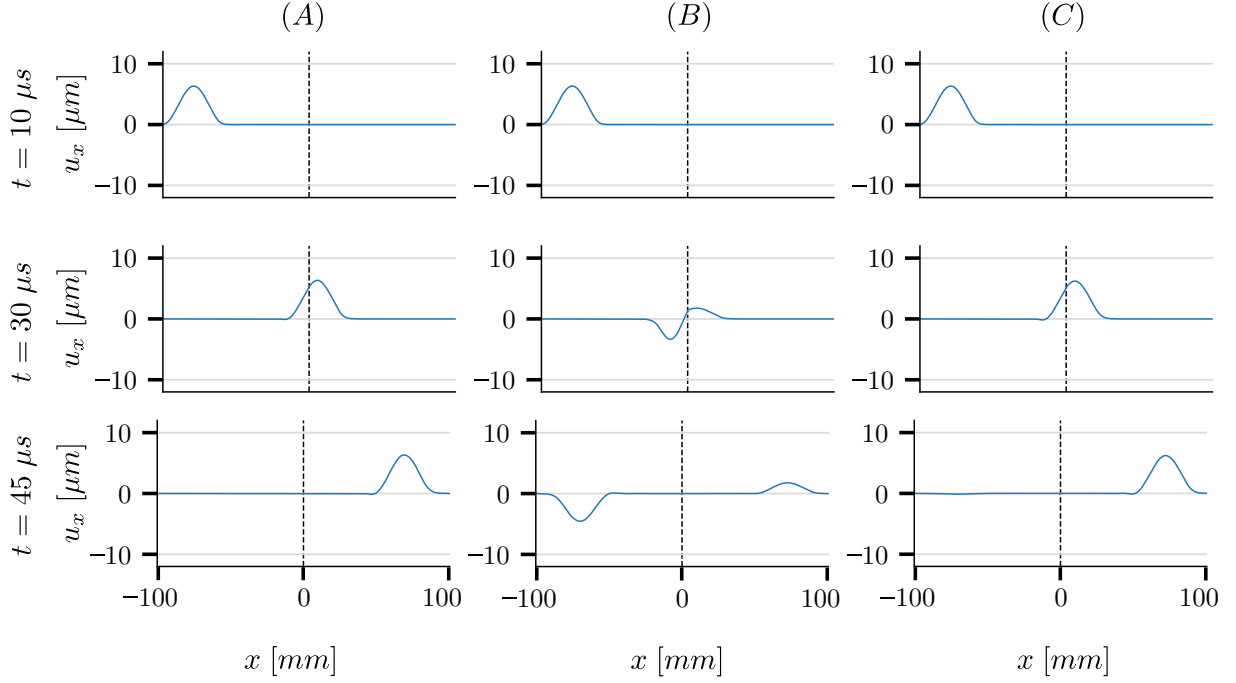

    \caption{Wave propagation through the bar for the cases A, B and C at different time steps.}
    \label{fig:spurious_reflections_2}
\end{figure}

The wave is visualised using the displacement field $u_x$, obtained at any point along the bar by averaging the displacements of all points in the $y$- and $z$-directions corresponding to the same $x$-location in the reference configuration. The results are shown in \cref{fig:spurious_reflections_2}. In case~(A), corresponding to uniform discretisation, the wave propagates smoothly without any reflection at the centre of the bar. In contrast, in case~(B), which employs a non-uniform discretisation and spatially varying horizon, a spurious reflection is observed at the centre of the bar at $30\mathrm{\mu s}$. This reflection is a non-physical artefact arising from the asymmetric interactions between the points in the left and right half-sections, which violate balance laws and introduce an effective impedance mismatch.

\begin{figure}[!htbp]
    \centering
    \def\svgwidth{0.65\textwidth}
    
    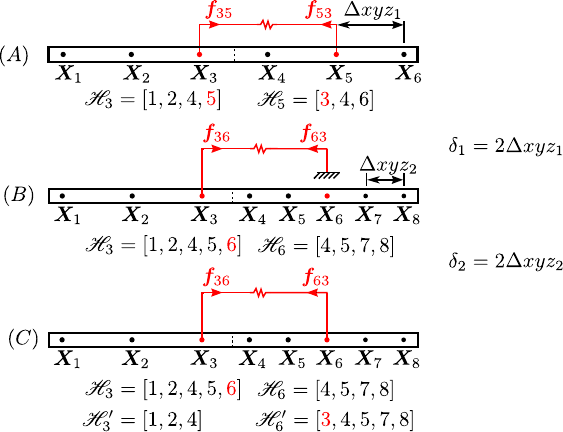

    \caption{Illustration of the bar discretisation with $\delta = 2 \Delta xyz$ for cases (A), (B) and (C). The internal interaction forces between two specific points is shown in red color. And the horizon and dual horizon is also mentioned.}
    \label{fig:homogeneous_and_non_homogeneous_diagrams}
\end{figure}

When a uniform discretisation and horizon are used, the forces in SHBB-PD are symmetric. For example, as shown in \cref{fig:homogeneous_and_non_homogeneous_diagrams} (A), the horizons of points $\vX_3$ and $\vX_5$ contain each other and hence the same equal and opposite forces $\vf_{35} = - \vf_{53}$ are exerted on each other, resulting in smooth wave propagation without any reflections. However, in the case of non-uniform discretisation and horizon, this symmetry is lost. For example in \cref{fig:homogeneous_and_non_homogeneous_diagrams} (B), the horizon of point $\vX_3$ contains $\vX_6$, whereas the horizon of $\vX_6$ does not contain $\vX_3$. Consequently, $\vX_3$ experiences a force $\vf_{36}$ without a corresponding reaction force $\vf_{63}$ acting on $\vX_6$. This asymmetric interaction violates balance laws and introduces an effective impedance mismatch, which leads to spurious wave reflections at the interface between the two regions. Physically, this can be interpreted as $\vX_3$ interacting with an artificial boundary, causing partial reflection of the wave, as observed in case~(B) of \cref{fig:spurious_reflections_2}.

In contrast, in case~(C), which employs the dual-horizon bond-based peridynamics formulation, the forces come not only from the horizon of the point but also from the dual-horizon. Hence, point $\vX_3$ experiences the force $\vf_{36}$ from its horizon, and point $\vX_6$ experiences the force $\vf_{63}$ from its dual-horizon, which is equal and opposite to $\vf_{36}$, i.e. $\vf_{63} = - \vf_{36}$. This results in symmetric interactions that satisfy balance laws and eliminate the effective impedance mismatch, thus mitigating spurious wave reflections at the interface between the two half-sections, as shown in case~(C) of \cref{fig:spurious_reflections_2}. It should be noted, however, that there is still a small reflection, although it is not visible in the figure, as it is significantly smaller than the reflection observed in case~(B). This minor effect is attributed to variations in bond micro-modulus across the interface and not to asymmetric interactions.

\begin{figure}[!htbp]
    \centering

    \includegraphics{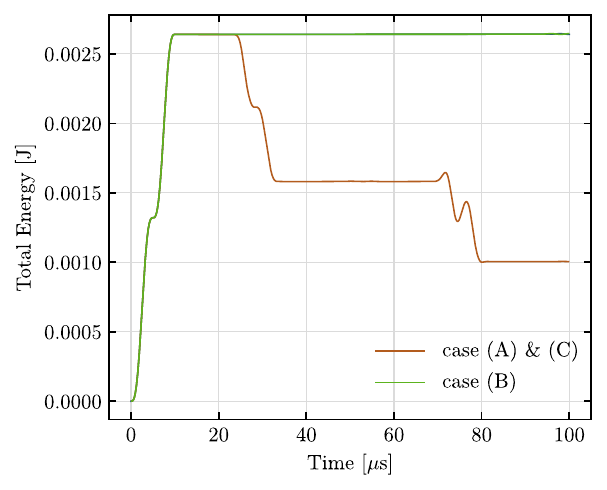}

    \caption{Total energy diagram for the 3 different wave propagation cases.}
    \label{fig:wave_energy_comparison}
\end{figure}

This behaviour is further reflected in the energy evolution shown in \cref{fig:wave_energy_comparison}, where the total energy of the system is plotted as a function of time for all three cases. In case~(A), the total energy remains constant throughout the simulation, indicating the absence of reflections or numerical artefacts. In case~(B), a drop in total energy is observed around $t = 30\,\mathrm{\mu s}$, corresponding to the onset of spurious reflections and indicating that part of the wave energy is transmitted to the artificial boundary while the rest is reflected or transmitted, leading to a decrease in the total energy of the system. In contrast, in case~(C), the total energy remains nearly constant, similar to case~(A), demonstrating that the dual-horizon bond-based peridynamics formulation effectively mitigates spurious reflections and preserves energy.

\section{Constitutive model} \label{sec:constitutive_model}

\subsection{Micro-elastic material}

To complete the formulation, we specify a constitutive material model that defines the micro-potential $\pi^{\textrm{int}}_{ij}$ and consequently the pairwise force function $\vf_{ij}$. The \textit{micro-elastic} model commonly used for linear elastic materials in bond-based peridynamics assumes a quadratic dependence on the bond stretch, which is defined as

\begin{equation}
    s_{ij} (\vx_{ij}) = \frac{\|\vx_{ij} \| - \| \vX_{ij} \|}{\| \vX_{ij} \|}.
\end{equation}

The micro-potential for the bond between $\vX_i$ and its neighbour $\vX_j$ is

\begin{equation} \label{eq:micro_potential_bbpd}
    \pi^{\textrm{int}}_{ij} = \frac{1}{2} c\,(\delta_j) \, s_{ij}^{\,2} (\vx_{ij}) \|\vX_{ij}\| ,
\end{equation}

where $c\,(\delta_j)$ is the \textit{micro-modulus} characterizing the stiffness of the bond between points $\vX_i$ and $\vX_j$. Thus, the resulting pairwise force function can be obtained by substituting the micro-potential in \cref{eq:pairwise_force_function_general}, i.e.,

\begin{equation} \label{eq:pairwise_force_function_bbpd}
    \vf_{ij}(\vx_{ij}) = c\,(\delta_j) \, s_{ij} (\vx_{ij}) \frac{\vx_{ij}}{\|\vx_{ij}\|}\,.
\end{equation}

The micro-modulus $c\,(\delta)$ must be calibrated to reproduce the correct macroscopic elastic behaviour. By equating the peridynamic strain energy density to the classical strain energy density under isotropic expansion \cite{sillingMeshfreeMethodBased2005}, one obtains

\begin{equation}
    c\,(\delta) = 
    \begin{cases}
        \dfrac{E}{\delta^2 A} & \text{for 1D} \\[8pt]
        \dfrac{6 E}{\pi \delta^4} & \text{for 3D, where } \nu = \dfrac{1}{4} \\[8pt]
        \dfrac{9 E}{2\pi \delta^3 z} & \text{for plane stress, where } \nu = \dfrac{1}{3} \\[8pt]
        \dfrac{24 E}{5 \pi \delta^3 z} & \text{for plane strain, where } \nu = \dfrac{1}{4}, \\[8pt]
    \end{cases}
\end{equation}

where $E$ is the Young's modulus, $\nu$ is the Poisson's ratio, $z$ is the thickness for 2D cases, and $A$ is the cross-sectional area for 1D cases. As mentioned above, bond-based PD constrains the Poisson's ratio to $\nu = 1/4$ for 3D and 2D plane strain conditions, and to $\nu = 1/3$ under 2D plane stress condition. One can note that the DHBB-PD micro-modulus function for the different conditions is half of the one for SHBB-PD.

Thus, the Lagrangian of the spatially discrete body formulated using the DHBB-PD formulation can be expressed as

\begin{equation}
\begin{aligned}
    L(\vx, \dot{\vx}, t) &= T(\dot{\vx}) - \Pi(\vx, t) \\
     &= \frac{1}{2} \sum_{i  =1}^{N} \varrho_i V_i \|\dot{\vx}_i\|^2
     - \sum_{i  =1}^{N} \sum_{j \in \sH_i}
     c\,(\delta_j) \, s_{ij}^2 \, \|\vX_{ij}\| \, V_j V_i \, -\Pi^{\textrm{ext}}(\vx, t).
\end{aligned}
\end{equation}

\subsection{Fracture model}

A key advantage of peridynamics is the natural incorporation of fracture through bond breakage. Fracture is typically modelled using a history-dependent damage function $\mu_{ij}(t) \in \left\{0, 1\right\}$, which modifies the pairwise force function to account for bond failure. The damage function takes the value of 1 when the bond is intact and 0 when the bond is broken. Once a bond is broken, it cannot heal, i.e. the damage is irreversible. The most common approach to modelling bond failure is to assume that a bond breaks when the bond stretch exceeds a critical value $s_c$. Thus, the damage function can be defined as 

\begin{equation}
    \mu_{ij} (t) = \begin{cases}
        1 & \text{if} \, \, s_{ij} (\vx_{ij}(t')) < s_c \quad \forall \, 0 \leq t' \leq t, \\
        0 & \text{otherwise,}
    \end{cases}
\end{equation}

Thus, the modified pairwise force function becomes

\begin{equation} \label{eq:pairwise_force_function_bbpd_failure}
    \vf_{ij} = c\,(\delta_j) \, \mu_{ij} \, s_{ij} \, \frac{\vx_{ij}}{\|\vx_{ij}\|}\,.
\end{equation}

The critical stretch $s_c$ is a material parameter that controls the onset of fracture. It can be computed by comparing the energy required to break all the bonds per unit fracture area to the critical energy release rate $G_{c}$ \cite{sillingMeshfreeMethodBased2005}. Expressions for the critical stretch $s_c$ corresponding to different modelling scenarios are summarised as

\begin{equation}
    s_c \, (\delta) = 
    \begin{cases}
        \sqrt{\dfrac{5 G_c}{6 E \delta}} & \quad \text{for 3D} \\[10pt]
        \sqrt{\dfrac{4 \pi G_c}{9 E \delta}} & \quad \text{for plane stress} \\[10pt]
        \sqrt{\dfrac{5 \pi G_c}{12 E \delta}} & \quad \text{for plane strain}
    \end{cases}
\end{equation}

Thus, the total damage at a material point $\vX_i$, denoted by 
$D_i(t) = D(\vX_i,t)$, is defined as
\begin{equation}
    D_i(t)
    =
    1
    -
    \dfrac{
        \sum\limits_{j \in \sH_i} \mu_{ij}(t) V_j
    }{
        \sum\limits_{j \in \sH_i} V_j
    } .
    \label{eq:point_damage}
\end{equation}

Therefore, $D_i(t)$ represents the volume-weighted fraction of broken bonds connected to the material point $\vX_i$. The local damage satisfies $0 \leq D_i(t) \leq 1$, where $D_i(t)=0$ means that no bonds connected to $\vX_i$ are broken, while $D_i(t)=1$ means that all bonds connected to $\vX_i$ are broken.

\section{Variational integrators for bond-based peridynamics} \label{sec:variational_integrators_for_bbpd}

In the previous sections, the equations of motion for a bond-based peridynamic body with varying horizons have been derived in continuous time. To numerically integrate these equations, a temporal discretisation is required. In this section, we develop variational integrators, which are a class of numerical integrators derived from a discrete version of the Lagrange–d'Alembert principle rather than by directly discretising the equations of motion. This approach ensures that key geometric properties of the continuous system, such as symplecticity and momentum conservation, are preserved in the discrete setting, leading to good long-term stability and accuracy. 

\subsection{Discrete variational mechanics} \label{sec:discrete_variational_mechanics}

As introduced in \cref{sec:bond_based_pd_with_variable_horizons}, based on the Lagrange–d'Alembert principle, the motion of the system satisfies

\begin{equation}
    \delta \mathcal{S}[\vx] + \int_{T_0}^{T_1} \delta \vx\T \vb^{\mathrm{ext}} \left( \vx, \Dot{\vx} \right) \mathrm{d}t = 0,
\end{equation}

Note that in the following sections, we assume that the external forces enter the system through the external forces $\vb^{\mathrm{ext}}$ only. The variational integrator framework discretises this principle directly rather than the resulting differential equations. To do so, the time interval $\left[T_0, T_1\right]$ is partitioned into discrete time points $t^k = T_0 + kh$, where $h$ denotes the uniform time step size, with $k = 0,1,\dots,N_t$. The continuous trajectory $\vx(t)$ is approximated by a discrete sequence of configurations $ \vx^0, \vx^1, \dots , \vx^{N_t}$, where $\vx^k = \vx(t^k)$.

The discrete Lagrangian $L_d$ approximates the action integral over a single time step, namely,

\begin{equation} \label{eq:discrete_lagrangian_expression}
    L_d(\vx^k, \vx^{k+1}) \approx \int_{t^k}^{t^{k+1}} L(\vx, \Dot{\vx}, t) \, \mathrm{d}t.
\end{equation}

Thus, the action of the system is approximated by the discrete action sum

\begin{equation} \label{eq:discrete_action_expression}
    \mathcal{S}_d [x_d] = \sum_{k=0}^{N_t-1} L_d(\vx^k, \vx^{k+1}),
\end{equation}

where $\vx_d = (\vx^0,\vx^0,\dots,\vx^{N_t})$ denote the complete discrete trajectory. In the discrete setting, the virtual work integral over each time step is approximated as 

\begin{equation}\label{eq:bext_eqn_integral_disc}
\int_{t^{k}}^{t^{k+1}}
(\delta \vx)^{\mathrm T} \,
\vb^{\mathrm{ext}}(\vx,\dot{\vx})
\,\mathrm{d}t
\approx
(\delta \vx^{k})^{\mathrm T} \,
\vb_{d}^{\mathrm{ext},-}(\vx^{k},\vx^{k+1})
+
(\delta \vx^{k+1})^{\mathrm T}
\vb_{d}^{\mathrm{ext},+}(\vx^{k},\vx^{k+1}),
\end{equation}

where $\vb^{\mathrm{ext,-}}_d$ and $\vb^{\mathrm{ext,+}}_d$ are the left and right discrete forces \cite{MarsdenWest2001}, respectively. Thus, the discrete Lagrange–d'Alembert principle reads

\begin{equation}
    \delta \sum_{k=0}^{N_t-1} L_d(\vx^k, \vx^{k+1})
    +
    \sum_{k=0}^{N_t-1}
    \left[
    (\delta \vx^{k})^{\mathrm T} \,
    \vb_{d}^{\mathrm{ext},-}(\vx^{k},\vx^{k+1})
    +
    (\delta \vx^{k+1})^{\mathrm T} \,
    \vb_{d}^{\mathrm{ext},+}(\vx^{k},\vx^{k+1})
    \right]
    =
    0 ,
\end{equation}

 $\forall \delta \vx^k, k = 1, \dots , N_t-1$, where $\delta \vx^0 = \delta \vx^{N_t} = 0$. As necessary conditions, this yields the \textit{forced discrete Euler–Lagrange equations}

\begin{equation} \label{eq:DEL_eqns}
    D_2 L_d(\vx^{k-1}, \vx^k) + D_1 L_d(\vx^k, \vx^{k+1}) + \vb^{\mathrm{ext,+}}_d \left( \vx^{k-1}, \vx^{k}\right) + \vb^{\mathrm{ext,-}}_d \left( \vx^{k}, \vx^{k+1} \right) = \mathbf{0},
\end{equation}

where $D_1 L_d$ and $D_2 L_d$ denote the partial derivatives of the discrete Lagrangian with respect to the first and second argument, respectively. Finally, since our Lagrangian is regular, we can use the discrete Legendre transforms to define the discrete momenta as

\begin{align} \label{eq:discrete__forced_legendre_transforms}
    \vp^{k,-} &= - D_1 L_d (\vx^k, \vx^{k+1}) - \vb^{\mathrm{ext,-}}_d (\vx^k, \vx^{k+1}), \\[4pt]
    \vp^{k,+} &= D_2 L_d (\vx^{k-1}, \vx^{k}) + \vb^{\mathrm{ext,+}}_d (\vx^{k-1}, \vx^{k}) \, .
\end{align}

Different quadrature rules for the approximations in \cref{eq:discrete_lagrangian_expression} and \cref{eq:bext_eqn_integral_disc} lead to integrators of varying accuracy. Using the trapezoidal rule in \cref{eq:discrete_lagrangian_expression}, we can write the expression for discrete action \cref{eq:discrete_action_expression}, as

\begin{equation}
    \begin{aligned}
        \mathcal{S}_d [\vx_d] = \sum_{k=0}^{N_t-1} \sum_{i=0}^{N} \left[ \dfrac12 \varrho_i V_i h \left\| \dfrac{\vx_i^{k+1} - \vx_i^{k}}{h} \right\|^2 - \sum_{j \in \sH_i} h \left( \dfrac{\pi^{\mathrm{int}}_{ij} \left( \vx_{ij}^{k+1}, \vX_{ij} \right) + \pi^{\mathrm{int}}_{ij} \left( \vx_{ij}^{k}, \vX_{ij} \right)}{2} \right) V_i V_j \right],
    \end{aligned}
\end{equation}

where $\vx_{ij}^k = \vx_j^k - \vx_i^k$ approximates the deformed bond vector $\vx_{ij} (t^k)$. Also, we can write the discrete forces at point $\vx_n$ as

\begin{align}
    \vb^{\mathrm{ext,-}}_{d,n} \left( \vx^{k}, \vx^{k+1} \right) &= \dfrac{h}{2} \, \vb_{n}^{k+1} , \\
    \vb^{\mathrm{ext,+}}_{d,n} \left( \vx^{k-1}, \vx^{k} \right) &= \dfrac{h}{2} \, \vb_{n}^{k} .
\end{align}

This quadrature choice leads to a second-order accurate explicit time integration scheme, known as the \textit{velocity-Verlet} (VV) scheme. Similarly, we can use the left-endpoint quadrature rule to obtain the first-order accurate explicit \textit{symplectic Euler} (SE) scheme. 

In this conservative case, variational integrators constructed in this manner are symplectic maps and inherit the geometric structure of the underlying Hamiltonian system. Consequently, they satisfy a discrete version of Noether’s theorem, leading to exact conservation of momenta associated with symmetries, and exhibit near-conservation of energy over long time intervals. These structure-preserving properties result in superior long-term stability and accuracy compared to standard time integration schemes \cite{MarsdenWest2001}. 

\subsection{Asynchronous variational integrators}

While standard variational integrators employ a single global time step, the presence of a spatially varying horizon and localised fracture phenomena motivate the use of different time step sizes in different regions of the domain. Asynchronous variational integrators and multirate integrators extend the discrete variational framework to accommodate such multi-rate time stepping, allowing different potentials to be evaluated at different times \cite{lewAsynchronousVariationalIntegrators2003,lewThesisCaltech2003,ober-blobaumVariationalMultirateIntegrators2024,leyendeckerVariationalApproachMultirate2013}.

\begin{figure}[!htbp]
    \centering
    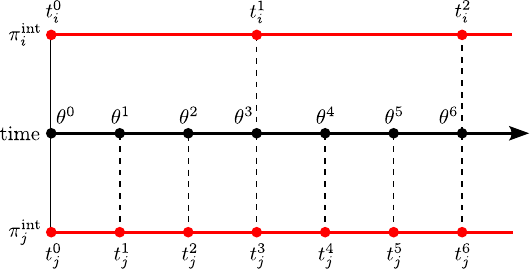
    \caption{Illustration of the discrete time grid for two potentials. Potential $\pi^{\mathrm{int}}_{i}$ is updated at a coarser local time grid, while $\pi^{\mathrm{int}}_{j}$ is updated at a finer local time grid. The nodes in the global time grid are the union of the nodes in the two local time grids. For the global time node $\theta^3:$ $\mathcal{K} (3) = \{i, j\}, \, \mathcal{J} (3, i) = 1, \, \mathcal{J} (3, j) = 3$ and $\mathcal{A}(3, i) = \sH_i, \mathcal{A}(3, j) = \sH_j$. }
    \label{fig:example_time_grid_for_two_potentials}
\end{figure}

A common practice is to assign an independent update time grid to each potential \cite{lewAsynchronousVariationalIntegrators2003}. In the present peridynamic setting, each potential $\pi^{\mathrm{int}}_i$, which includes all the potentials as described in $\cref{eq:pi_i_equations}$ is assigned a local sequence of update time nodes $\mathcal{T}_i = \{ t_i^0 , \dots , t_i^{M_i} \}$, with $t_i^0 = T_0 < \dots < t_i^{M_i} = T_1$ such that all the bond potentials $\pi^{\mathrm{int}}_{i}$ are updated at the times in this sequence. All update time nodes are then merged into a single strictly increasing global time grid denoted as

\begin{equation}
    \theta = \bigcup_{i} \mathcal{T}_i = \left\{ \theta^0 , \dots , \theta^M \right\},
\end{equation}

where $M$ is the total number of distinct time nodes in the global time grid, and $\theta^0 < \theta^1 < \dots < \theta^M$ is assumed. The configuration of the system at time node $\theta^k$ is denoted by $\vx^k$. For each global time node $\theta^k$, define the \textbf{active potential set} $\mathcal{K} (k)$ 

\begin{equation}
    \mathcal{K} (k) = \left\{ i \middle| \; \exists \, \tau, t_i^{\tau} = \theta^k \right\},
\end{equation}

that is, the set of potentials that are updated at time \(\theta^k\). To relate the global and local time nodes, a mapping between the two index sets is introduced. The function $ \mathcal{J} (k, i)$ returns the local time index $\tau$ corresponding to a global update time $\theta^k$, i.e.

\begin{equation}
    \mathcal{J} (k, i) = \tau, t_i^{\tau} = \theta^k .
\end{equation}

The discrete Lagrangian, and consequently the action, in the asynchronous formulation is constructed by aggregating contributions from individual interactions over their respective local time intervals. Depending on the quadrature rule used for approximating the discrete Lagrangian, different asynchronous variational integrators can be obtained. In the following sections, first- and second-order accurate AVI schemes are derived based on the left-endpoint rule and the trapezoidal rule, respectively.

\subsubsection{Asynchronous velocity-Verlet (AVV)} \label{sec:async_velocity_verlet}

Using a trapezoidal quadrature rule for the potential energy leads to a second-order accurate asynchronous velocity-Verlet scheme. The resulting discrete action consists of kinetic contributions evaluated on the global time nodes and potential contributions evaluated on local time nodes associated with active interactions. The resulting discrete action can be written as

\begin{equation}
    \begin{aligned}
        \mathcal{S}_d [\vx_d] = & \sum_{k=0}^{M-1} \left( \theta^{k+1} - \theta^k \right) T \left( \dfrac{\vx^{k+1} - \vx^{k}}{\theta^{k+1} - \theta^k} \right) - \sum_{i=1}^{N} \sum_{\tau=0}^{M_k-1} \left[ t_i^{\tau+1} - t_i^{\tau} \right] \left( \dfrac{\pi^{\mathrm{int}}_i \left( \vx (t_i^{\tau+1}) \right) + \pi^{\mathrm{int}}_i \left( \vx (t_i^{\tau}) \right)}{2} \right) V_i,
    \end{aligned}
\end{equation}

where $\pi^{\mathrm{int}}_i \left( \vx (t_i^{\tau}) \right)$ denotes the internal energy density of $\sH_i$ at the local time $t_i^{\tau}$. Expanding the kinetic and potential energy contributions, we can write the discrete action as

\begin{equation} \label{eq:avv_discrete_action}
    \begin{aligned}
        \mathcal{S}_d [\vx_d] = & \sum_{k=0}^{M-1} \sum_{i=1}^{N} \dfrac12 \varrho_i V_i \left( \theta^{k+1} - \theta^k \right) \left\| \dfrac{\vx_i^{k+1} - \vx_i^{k}}{\theta^{k+1} - \theta^k} \right\|^2 - \\
        & \sum_{i=1}^{N} \sum_{\tau=0}^{M_k-1} \left[ t_i^{\tau+1} - t_i^{\tau} \right] \sum_{j \in \sH_i} \left[ \dfrac{\pi_{ij}^{\mathrm{int}}(\vx_{ij}(t_i^{\tau+1}), \vX_{ij}) + \pi_{ij}^{\mathrm{int}}(\vx_{ij}(t_i^{\tau}), \vX_{ij}) }{2} \right] V_i V_j
    \end{aligned}
\end{equation}

\begin{figure}[!htbp]
    \centering
    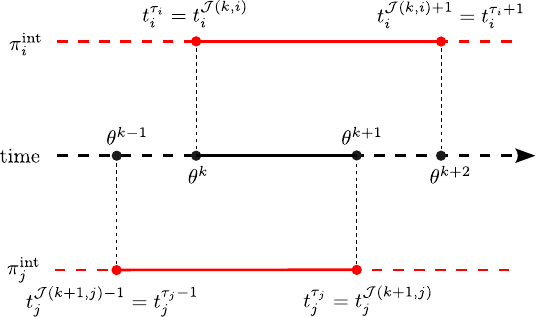
    \caption{Schematic illustration of the construction of the discrete Lagrangian \cref{eq:avv_discrete_lagrangian} over a global time interval $[\theta^k, \theta^{k+1}]$ in the asynchronous variational integrator framework. The kinetic energy is evaluated on the global time grid, while the potential energy contributions are evaluated on local time grids. The total potential contribution over the interval is assembled from contributions of the interactions active at $\theta^k$ and $\theta^{k+1}$.}
    \label{fig:two_point_potential_diagram}
\end{figure}

In order to state the discrete equations of motion, resulting from the discrete Lagrange-d'Alembert principle, we introduce the notion of the \textbf{active family set} $\mathcal{A} (k, i)$ and the \textbf{active dual-family set} $\mathcal{A'} (k, i)$. For a given potential $i \in \mathcal{K} (k)$, the active family set $\mathcal{A} (k, i)$ contains the indices of all the interactions associated with the $i^{\mathrm{th}}$ potential, i.e. $\mathscr{H}_i$. Whereas, the active dual-family set $\mathcal{A'} (k, i)$ contains indices of all the interactions associated with the active potentials at $\theta^k$ that involve point $i$ as a neighbour. More precisely, they can be defined as

\begin{equation}
    \mathcal{A}(k,i)
    =
    \begin{cases}
    \mathscr{H}_i, & i\in\mathcal{K}(k),\\[0.3em]
    \varnothing, & \text{otherwise}
    \end{cases} \hspace{2cm}
    \mathcal{A}'(k,i) = \left\{j\in\mathscr{H}_i' \;\middle|\; j\in\mathcal{K}(k) \right\} .
\end{equation}

Thus, we can obtain the discrete Euler-Lagrange equations for each point by making the discrete action stationary with respect to variations of the trajectory, leading to the following update equations for the positions and velocities of the points at each global time node,

\begin{equation} \label{eq:AVV_scheme}
    \begin{aligned}
    \vv_i^{k+\frac{1}{2}}
    &=
    \vv_i^k
    + \dfrac{\theta^{k+1} - \theta^k}{2\varrho_i}
        \latBoldUp{b}^{\mathrm{ext},k}_i
    \begin{aligned}[t]
    \,+\,\dfrac{t_i^{\mathcal{J}(k,i)+1} - t_i^{\mathcal{J}(k,i)}}{2\varrho_i}
    &\sum_{j\in\mathcal{A}(k, i)}
        \dfrac{\partial \pi_{ij}^{\mathrm{int}}}{\partial \vx_{ij}^{k}}
        \left(\vx_{ij}^{k}, \vX_{ij}\right) V_j
    \\
    -&\sum_{j\in\mathcal{A}'(k, i)}
        \dfrac{t_j^{\mathcal{J}(k,j)+1} - t_j^{\mathcal{J}(k,j)}}{2\varrho_i}
        \dfrac{\partial \pi_{ji}^{\mathrm{int}}}{\partial \vx_{ji}^{k}}
        \left(\vx_{ji}^{k}, \vX_{ji}\right) V_j
    \end{aligned}
    \\[0.75em]
    \vx_i^{k+1}
    &=
    \vx_i^k
    +
    (\theta^{k+1}-\theta^k)\,\vv_i^{k+\frac12}
    \\[0.75em]
    \vv_i^{k+1}
    &=
    \vv_i^{k+\frac{1}{2}}
    + \dfrac{\theta^{k+1} - \theta^k}{2\varrho_i}
        \latBoldUp{b}^{\mathrm{ext},k+1}_i
    \begin{aligned}[t]
    &\;+\dfrac{t_i^{\mathcal{J}(k+1,i)} - t_i^{\mathcal{J}(k+1,i)-1}}{2\varrho_i}
    \sum_{j\in\mathcal{A}(k+1, i)}
        \dfrac{\partial \pi_{ij}^{\mathrm{int}}}{\partial \vx_{ij}^{k+1}}
        \left(\vx_{ij}^{k+1}, \vX_{ij}\right) V_j
    \\
    &\;-\sum_{j\in\mathcal{A}'(k+1, i)}
        \dfrac{t_j^{\mathcal{J}(k+1,j)} - t_j^{\mathcal{J}(k+1,j)-1}}{2\varrho_i}
        \dfrac{\partial \pi_{ji}^{\mathrm{int}}}{\partial \vx_{ji}^{k+1}}
        \left(\vx_{ji}^{k+1}, \vX_{ji}\right) V_j
    \end{aligned}
    \end{aligned}
\end{equation}

The velocities in \cref{eq:AVV_scheme} are obtained from the discrete conjugate momenta introduced through the discrete Legendre transforms in \cref{eq:discrete__forced_legendre_transforms}. The scheme is initialised from the prescribed initial positions and velocities of the material points, after which the staggered velocity updates are applied recursively at the global time nodes. In this form, the method can be interpreted as a multirate extension of the classical velocity--Verlet integrator. It preserves the staggered velocity structure and inherits the associated second-order accuracy in time. Furthermore, since the update equations are derived from a discrete variational principle, the scheme retains the symplectic structure and exhibits favourable long-term energy behaviour. The asynchronous formulation also allows interactions associated with different horizon sizes to be evaluated at their respective stable time steps, thereby improving computational efficiency for non-uniform peridynamic discretisations.

Following the construction illustrated in \cref{fig:two_point_potential_diagram}, the discrete Lagrangian associated with the global time interval $\left[\theta^k, \theta^{k+1}\right]$ can also be explicitly written as

\begin{equation} \label{eq:avv_discrete_lagrangian}
    \begin{aligned}
        L_d (\vx^k, \vx^{k+1}) = & \sum_{i=1}^{N} \dfrac12 \varrho_i V_i \left( \theta^{k+1} - \theta^k \right) \left\| \dfrac{\vx_i^{k+1} - \vx_i^{k}}{\theta^{k+1} - \theta^k} \right\|^2 - \\
        & \sum_{i \in \mathcal{K}(k)} \left[ \dfrac{t_i^{\mathcal{J}(k, i)+1} - t_i^{\mathcal{J}(k, i)}}{2} \right] \sum_{j \in \sH_i} \pi^{\mathrm{int}}_{ij} \left( \vx^k_{ij}, \vX_{ij} \right) V_i V_j - \\
        & \sum_{i \in \mathcal{K}(k+1)} \left[ \dfrac{t_i^{\mathcal{J}(k+1, i)} - t_i^{\mathcal{J}(k+1, i)-1}}{2} \right] \sum_{j \in \sH_i} \pi^{\mathrm{int}}_{ij} \left( \vx^{k+1}_{ij}, \vX_{ij} \right) V_i V_j
    \end{aligned}
\end{equation}

It is straightforward to show that the above VI scheme \cref{eq:AVV_scheme} is nothing but \cref{eq:DEL_eqns} for the discrete Lagrangian \cref{eq:avv_discrete_lagrangian}. The algorithmic routing of the AVV scheme is summarised in \cref{alg:avv_algorithm}. The algorithm uses a priority queue (PQ) to manage the global time nodes, ensuring that the interactions are updated in the correct order. The forces are accumulated over the local time intervals, and the velocities and positions are updated accordingly at each global time node.

\begin{algorithm}
\caption{Asynchronous velocity-Verlet (AVV) integration Scheme}
\label{alg:avv_algorithm}
\begin{algorithmic}[1]

\State \textbf{Input:} $\vx^0, \vv^0, \theta^0, T_0, T_1$, set of $t^{\tau}_i$ for all $i = 1,\dots, N$
\State \textbf{Initialization:}
\State $k = 0, \vv = \vv^0, \vx = \vx^k, \theta^{\mathrm{old}} = \theta^0$
\State $\vF^+ = \vF^- = \boldsymbol{0}$
\For{$i = 1,\dots, N$}
    \State Push $(t_i^0, i)$ into \textit{PQ}
\EndFor
\State \textbf{Integration between } $T_0$ \textbf{and} $T_1$:
\While{\textit{PQ} is not empty}
    \State extract next element: pop $(t^{\tau}_i, i)$ from \textit{PQ}
    \State $\theta^{\mathrm{new}} = t^{\tau}_i$
    \If{$\theta^{\mathrm{new}} > \theta^{\mathrm{old}}$}
        \For{$i = 1,\dots, N$}
            \State $\vv_i = \vv_i + \dfrac12 \dfrac{\vF_i^-}{\varrho_i} + \dfrac{\theta^{\mathrm{new}} - \theta^{\mathrm{old}}}{2 \varrho_i} \, \vb_i^{\mathrm{ext}} (\theta^{\mathrm{old}})$
        \EndFor
        \State $\vx^k = \vx, \vv^k = \vv, \theta^k = \theta^{\mathrm{old}}$
        \State $k = k+ 1$
        \For{$i = 1,\dots, N$}
            \State $\vv_i = \vv_i + \dfrac12 \dfrac{\vF_i^+}{\varrho_i} + \dfrac{\theta^{\mathrm{new}} - \theta^{\mathrm{old}}}{2 \varrho_i} \, \vb_i^{\mathrm{ext}} (\theta^{\mathrm{old}})$
            \State $\vx_i = \vx_i + \left( \theta^{\mathrm{new}} - \theta^{\mathrm{old}} \right) \vv_i$
        \EndFor
        \State $\vF^+ = \vF^- = \boldsymbol{0}$
    \EndIf
    \State $\theta^{\mathrm{new}} = \theta^{\mathrm{old}}$
    \If{$\tau > 0$}
        \For{all $j \in \mathcal{A}(k, i)$}
            \State $\begin{aligned}
                \vF_i^- &= \vF_i^- + (t_i^{\tau} - t_i^{\tau - 1}) \sum_{j}
                \dfrac{\partial \pi^{\mathrm{int}}_{ij}}{\partial \vx^k_{ij}}
                \left( \vx^{k}_{ij}, \vX_{ij} \right)V_j
            \end{aligned}$
            \State $\begin{aligned}
                \vF_j^- &= \vF_j^- - (t_i^{\tau} - t_i^{\tau - 1})
                \dfrac{\partial \pi^{\mathrm{int}}_{ij}}{\partial \vx^k_{ij}}
                \left( \vx^{k}_{ij}, \vX_{ij} \right)V_j
            \end{aligned}$
        \EndFor
    \EndIf

    \If{$\tau < M_i$}
        \For{all $j \in \mathcal{A}(k, i)$}
            \State $\begin{aligned}
                \vF_i^+ &= \vF_i^+ + (t_i^{\tau + 1} - t_i^{\tau}) \sum_{j}
                \dfrac{\partial \pi^{\mathrm{int}}_{ij}}{\partial \vx^k_{ij}}
                \left( \vx^{k}_{ij}, \vX_{ij} \right)V_j
            \end{aligned}$
            \State $\begin{aligned}
                \vF_j^+ &= \vF_j^+ - (t_i^{\tau + 1} - t_i^{\tau})
                \dfrac{\partial \pi^{\mathrm{int}}_{ij}}{\partial \vx^k_{ij}}
                \left( \vx^{k}_{ij}, \vX_{ij} \right)V_j
            \end{aligned}$
        \EndFor
        \State Push $(t_i^{\tau + 1},i)$ into \textit{PQ}
    \EndIf
\EndWhile
\For{$i = 1,\dots, N$}
    \State $\vv_i = \vv_i + \dfrac12 \dfrac{\vF_i^-}{\varrho_i} + \dfrac{\theta^{\mathrm{new}} - \theta^{\mathrm{old}}}{2 \varrho_i} \, \vb_i^{\mathrm{ext}} (\theta^{\mathrm{new}})$
\EndFor
\State $\vx^k = \vx, \vv^k = \vv, \theta^k = \theta^{\mathrm{old}}$

\end{algorithmic}
\end{algorithm}

\subsubsection{Asynchronous symplectic-Euler (ASE)}

Similar to the above AVV scheme, we can derive a first-order accurate asynchronous symplectic Euler scheme by using the left-endpoint quadrature rule for the potential energy. The resulting discrete action can be written as

\begin{equation}
    \begin{aligned}
        \mathcal{S}_d [\vx_d] = & \sum_{k=0}^{M-1} \left( \theta^{k+1} - \theta^k \right) T \left( \dfrac{\vx^{k+1} - \vx^{k}}{\theta^{k+1} - \theta^k} \right) - \sum_{i=1}^{N} \sum_{\tau=0}^{M_k-1} \left[ t_i^{\tau+1} - t_i^{\tau} \right] \pi^{\mathrm{int}}_i \left( \vx (t_i^{\tau}) \right) V_i,
    \end{aligned}
\end{equation}

Substituting the expression for the kinetic and potential energy contributions, we can write the discrete action as

\begin{equation}
    \begin{aligned}
        \mathcal{S}_d [\vx_d] = & \sum_{k=0}^{M-1} \sum_{i=1}^{N}  \dfrac12 \varrho_i V_i \left( \theta^{k+1} - \theta^k \right) \left\| \dfrac{\vx_i^{k+1} - \vx_i^{k}}{\theta^{k+1} - \theta^k} \right\|^2 - \sum_{i=1}^{N} \sum_{\tau=0}^{M_k-1} \left[ t_i^{\tau+1} - t_i^{\tau} \right] \sum_{j \in \sH_i} \pi_{ij}^{\mathrm{int}}(\vx_{ij}(t_i^{\tau}), \vX_{ij}) V_i V_j
    \end{aligned}
\end{equation}

Thus, the corresponding discrete Euler-Lagrange equations obtained by making the above discrete action stationary with respect to variations of the trajectory are given by the following update equations for the positions and velocities of the points at each global time nodes

\begin{equation} \label{eq:ASE_scheme}
    \begin{aligned}
    \vv_i^{k+1}
    &=
    \vv_i^k
    +
    \frac{\theta^{k+1}-\theta^k}{\varrho_i}\,\vb_i^{\mathrm{ext},k}
    +
    \frac{t_i^{\mathcal J(k,i)+1}-t_i^{\mathcal J(k,i)}}{\varrho_i}
    \sum_{j\in\mathcal{A}(k,i)}
    \frac{\partial \pi_{ij}^{\mathrm{int}}}{\partial \vx_{ij}^{k}}
    \left(\vx_{ij}^{k},\vX_{ij}\right)V_j
    \\
    &\phantom{= \vv_i^k
    +\frac{\theta^{k+1}-\theta^k}{\varrho_i}\,\vb_i^{\mathrm{ext},k}}
    -
    \sum_{j\in\mathcal{A}'(k,i)}
    \dfrac{t_j^{\mathcal J(k,j)+1}-t_j^{\mathcal J(k,j)}}{\varrho_i}
    \frac{\partial \pi_{ji}^{\mathrm{int}}}{\partial \vx_{ji}^{k}}
    \left(\vx_{ji}^{k},\vX_{ji}\right)V_j,
    \\[0.75em]
    \vx_i^{k+1}
    &=
    \vx_i^k
    +
    \left(\theta^{k+1}-\theta^k\right)\vv_i^{k+1}.
    \end{aligned}
\end{equation}

Similarly, we can write the discrete Lagrangian for the above ASE scheme as

\begin{equation}
    \begin{aligned}
        L_d (\vx^k, \vx^{k+1}) = & \sum_{i=1}^{N} \dfrac12 \varrho_i V_i \left( \theta^{k+1} - \theta^k \right) \left\| \dfrac{\vx_i^{k+1} - \vx_i^{k}}{\theta^{k+1} - \theta^k} \right\|^2 - \\
        & \sum_{i \in \mathcal{K}(k)} \left[ t_i^{\mathcal{J}(k, i)+1} - t_i^{\mathcal{J}(k, i)} \right] \sum_{j \in \sH_i} \pi^{\mathrm{int}}_{ij} \left( \vx^k_{ij}, \vX_{ij} \right) V_i V_j
    \end{aligned}
\end{equation}

Similar to the AVV scheme, the algorithmic routine of ASE scheme is given in \cref{alg:ase_algorithm}.

\begin{algorithm}
\caption{Asynchronous symplectic Euler (ASE) integration Scheme}
\label{alg:ase_algorithm}
\begin{algorithmic}[1]

\State \textbf{Input:} $\vx^0, \vv^0, \theta^0, T_0, T_f$, set of $t^{\tau}_i$ for all $i = 1,\dots, N$
\State \textbf{Initialization:}
\State $k = 0, \vv = \vv^0, \vx = \vx^k, \theta^{\mathrm{old}} = \theta^0$
\State $\vF^+ = \vF^- = \boldsymbol{0}$
\For{$i = 1,\dots, N$}
    \State Push $(t_i^0, i)$ into \textit{PQ}
\EndFor
\State \textbf{Integration between } $T_0$ \textbf{and} $T_1$:
\While{\textit{PQ} is not empty}
    \State extract next element: pop $(t^{\tau}_i, i)$ from \textit{PQ}
    \State $\theta^{\mathrm{new}} = t^{\tau}_i$
    \If{$\tau < M_k$}
        \For{all $j \in \mathcal{A}(k, i)$}
            \State $\begin{aligned}
                \vF_i &= \vF_i + (t_i^{\tau + 1} - t_i^{\tau}) \sum_{j}
                \dfrac{\partial \pi^{\mathrm{int}}_{ij}}{\partial \vx^k_{ij}}
                \left( \vx^{k}_{ij}, \vX_{ij} \right)V_j
            \end{aligned}$
            \State $\begin{aligned}
                \vF_j &= \vF_j - (t_i^{\tau + 1} - t_i^{\tau})
                \dfrac{\partial \pi^{\mathrm{int}}_{ij}}{\partial \vx^k_{ij}}
                \left( \vx^{k}_{ij}, \vX_{ij} \right)V_j
            \end{aligned}$
        \EndFor
        \State Push $(t_i^{\tau + 1},i)$ into \textit{PQ}
    \EndIf
    \If{$\theta^{\mathrm{new}} > \theta^{\mathrm{old}}$}
        \For{$i = 1,\dots, N$}
            \State $\vv_i = \vv_i + \dfrac{\vF_i}{\varrho_i} + \dfrac{\theta^{\mathrm{new}} - \theta^{\mathrm{old}}}{\varrho_i} \, \vb_i^{\mathrm{ext}} (\theta^{\mathrm{old}})$ \vspace{0.5mm}
            \State $\vx_i = \vx_i + \left( \theta^{\mathrm{new}} - \theta^{\mathrm{old}} \right) \vv_i$
        \EndFor
        \State $\vx^k = \vx, \vv^k = \vv, \theta^k = \theta^{\mathrm{old}}$
        \State $k = k+ 1$
        \State $\vF = \boldsymbol{0}$
    \EndIf
    \State $\theta^{\mathrm{new}} = \theta^{\mathrm{old}}$
\EndWhile

\end{algorithmic}
\end{algorithm}

\subsection{Numerical examples} \label{sec:numerical_examples}

In this section, we present a series of numerical examples to assess the performance of the proposed asynchronous variational integrators for bond-based peridynamics with non-uniform discretisations and spatially varying horizons. Two different problems are considered: (i) dynamic fracture of a pre-cracked plate under tension (also commonly referred to as the boundary tension test), and (ii) the Kalthoff–Winkler impact experiment.

These examples focus on dynamic fracture and are used to evaluate the performance of the asynchronous variational integrators. In particular, these examples illustrate the capability of the proposed approach to accommodate spatially varying time step sizes while preserving accuracy, stability, and physically consistent crack propagation. All simulations are performed using an in-house peridynamics code implemented in Python, with computationally intensive routines accelerated through compiled kernels.

\subsection{Critical-time step}

Before presenting the numerical examples, we briefly discuss the critical time step for explicit time integration schemes in peridynamics. The critical time step is an important parameter for ensuring the stability of the numerical solution. Several criteria exist to estimate the critical time step for explicit time integration schemes in peridynamics \cite{littlewoodEstimationCriticalTime2014}, and one common approach is based on the von Neumann stability analysis \cite{sillingMeshfreeMethodBased2005}, which approximates the critical time step $h_c$ as

\begin{equation}
    h_c = \min\limits_{i}
    \sqrt{\dfrac{2 \varrho_i}{
    \sum\limits_{j \in \sH_i} 
    \dfrac{c\,(\delta_j) V_j}{\| \vX_{ij} \|}
    }} \, .
\end{equation}

Usually a safety factor, $\alpha$ smaller than 1 is applied to ensure stability during the simulation. Thus, the time step used in the simulation is given by 

\begin{equation}
    h = \alpha \, h_c \,.
\end{equation}

In order to determine the critical time steps for different regions in the domain for the asynchronous time integration, one can compute the critical time step for each region by taking a minimum of the critical time steps of the points in that region. For example, for a region $\Omega_1$ containing points $i \in \Omega_1$, the critical time step can be estimated as

\begin{equation}
    h_{c1} = \min\limits_{i \in \Omega_1}
    \sqrt{\dfrac{2 \varrho_i}{
    \sum\limits_{j \in \sH_i} 
    \dfrac{c\,(\delta_j) V_j}{\| \vX_{ij} \|}
    }} \, .
\end{equation}

Thus, with AVI we use a common safety factor and multiply it with the critical time step of each region to determine the time step size for that region, i.e. $h_1 = \alpha h_{c1}$ for region $\Omega_1$.

\subsection{Pre-cracked plate under tension}

In order to study the effectiveness of the proposed asynchronous variational integrators in dynamic fracture simulations, we consider the classical benchmark problem of a rectangular pre-cracked plate subjected to uniaxial tension. This problem has been extensively studied in the peridynamic literature \cite{dipasqualeCrackPropagationAdaptive2014,rakiciDiscreteSurfaceCorrection2023}. 

In order to evaluate the performance of the AVI scheme relative to the standard velocity-Verlet (VV) integrator, two discretisation strategies are investigated. The first employs a uniform discretisation, while the second utilizes a non-uniform discretisation in which the mesh is locally refined along a prescribed path that approximates the expected crack trajectory. The spacing of the points in the refined region is chosen to be half of that in the coarse region. The refinement path is selected based on prior knowledge of the problem and represents the anticipated direction of crack propagation under local refinement. However, due to the nonlocal nature of the peridynamic formulation, the resulting crack path is not uniquely determined by this prescribed refinement and may differ from the trajectory obtained while using a uniformly fine discretisation.

This allows for a direct comparison of the fracture patterns obtained using the VV and AVV schemes under both uniform and non-uniform discretisations. In case of AVV, the time step size is determined based on the critical time step of each region, allowing for smaller time steps in the refined region and larger time steps in the coarser regions. In contrast, the VV scheme requires a uniform time step size across the entire domain, which must be chosen based on the smallest critical time step corresponding to the refined region.

\begin{figure}[!htbp]
    \centering

    \includegraphics{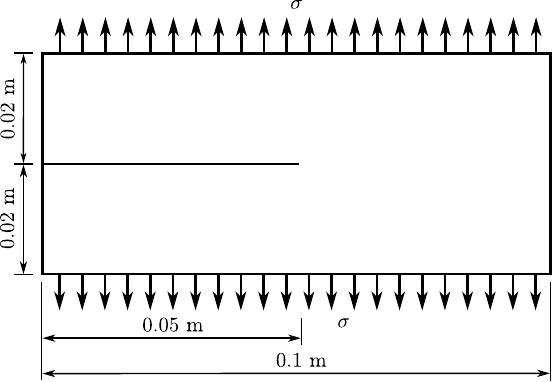}

    \caption{Setup of the pre-cracked plate under tension.}
    \label{fig:btt_diagram}
\end{figure}

The plate has dimensions of $0.1\,\mathrm{m} \times 0.04\,\mathrm{m}$ and contains a horizontal precrack of length $0.05\,\mathrm{m}$ starting from the left edge of the plate as shown in \cref{fig:btt_diagram}. The material is modelled as Soda-Lime glass with density $\varrho = 2440\mathrm{kg/m^3}$, and Young's modulus $E = 72\,\mathrm{GPa}$, and a fracture energy of $G_c = 135\,\mathrm{J/m^2}$. The simulations are performed in two dimensions under plane stress conditions. Accordingly, the bond-based peridynamic formulation employs the plane-stress micro-modulus, which restricts the Poisson’s ratio to $\nu = \frac13$. 

A constant tensile stress of $12 \mathrm{MPa}$ is applied on the top and bottom boundaries of the plate. This loading is implemented by converting the prescribed traction into equivalent body forces acting on the outermost rows of material points. The simulations are carried out for a total duration of $50 \mathrm{\mu s}$. In both the coarse and refined regions, the peridynamic horizon is chosen proportional to the local point spacing, with $ \delta = 3.015 \Delta xy$. Additionally, a uniform safety factor of $0.2$ is employed across all regions.

\begin{figure}[!htbp]
    \centering

    \includegraphics{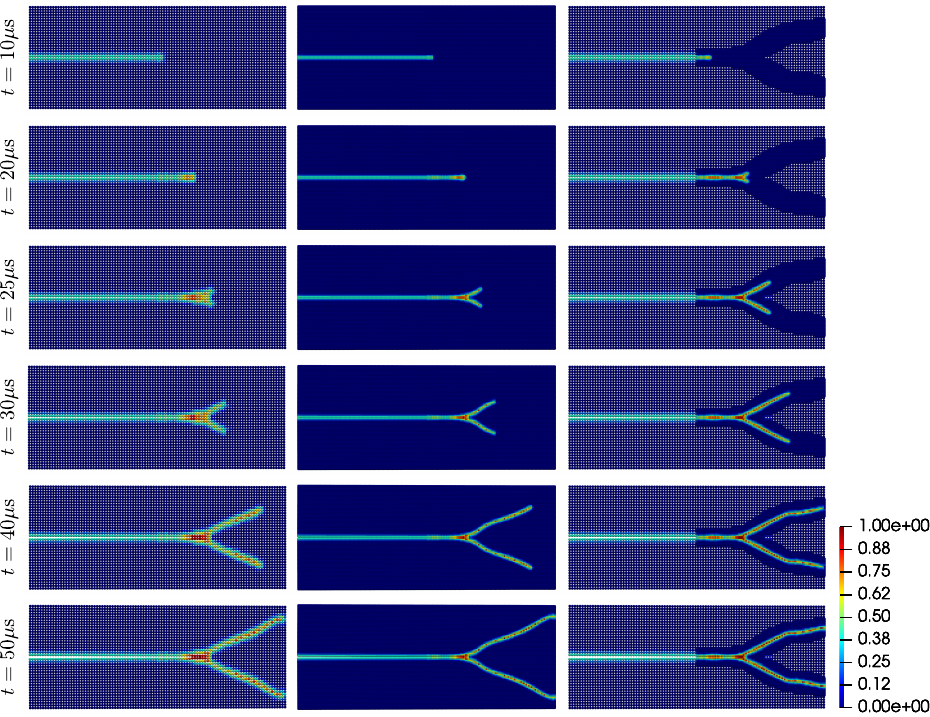}

    \caption{Temporal evolution of fracture patterns in a pre-cracked plate under tension simulated using the velocity-Verlet scheme with DHBB-PD formulation for three discretisation strategies: (left) a uniform discretisation with $100 \times 40$ points, (middle) a uniform discretisation with $200 \times 80$ points, and (right) a non-uniform discretisation with $5842$ points in total, comprising $3386$ points in the coarse region and $2456$ points in the refined region. In the non-uniform case, the coarse region uses the same spatial resolution as the $100 \times 40$ uniform discretisation, while the refined region uses the same spatial resolution as the $200 \times 80$ uniform discretisation. The refined region is placed along the expected crack path. In all cases, the horizon is chosen as $\delta = 3.015\,\Delta xy$ within each respective region.}
    \label{fig:btt_fracture_patterns}
\end{figure}

\cref{fig:btt_fracture_patterns} compares the fracture patterns obtained using uniform and non-uniform discretisations simulated using the VV scheme with DHBB-PD formulation. The results highlight the robustness of the dual-horizon formulation, which successfully handles non-uniform discretisations with spatially varying horizons while preserving a physically consistent crack path. In particular, the crack propagates along the expected horizontal trajectory even in the presence of local refinement, indicating that the dual-horizon approach preserves the correct transmission of stress waves across regions with different resolutions.

\begin{figure}[!htbp]
    \centering

    \includegraphics[width=0.8\textwidth]{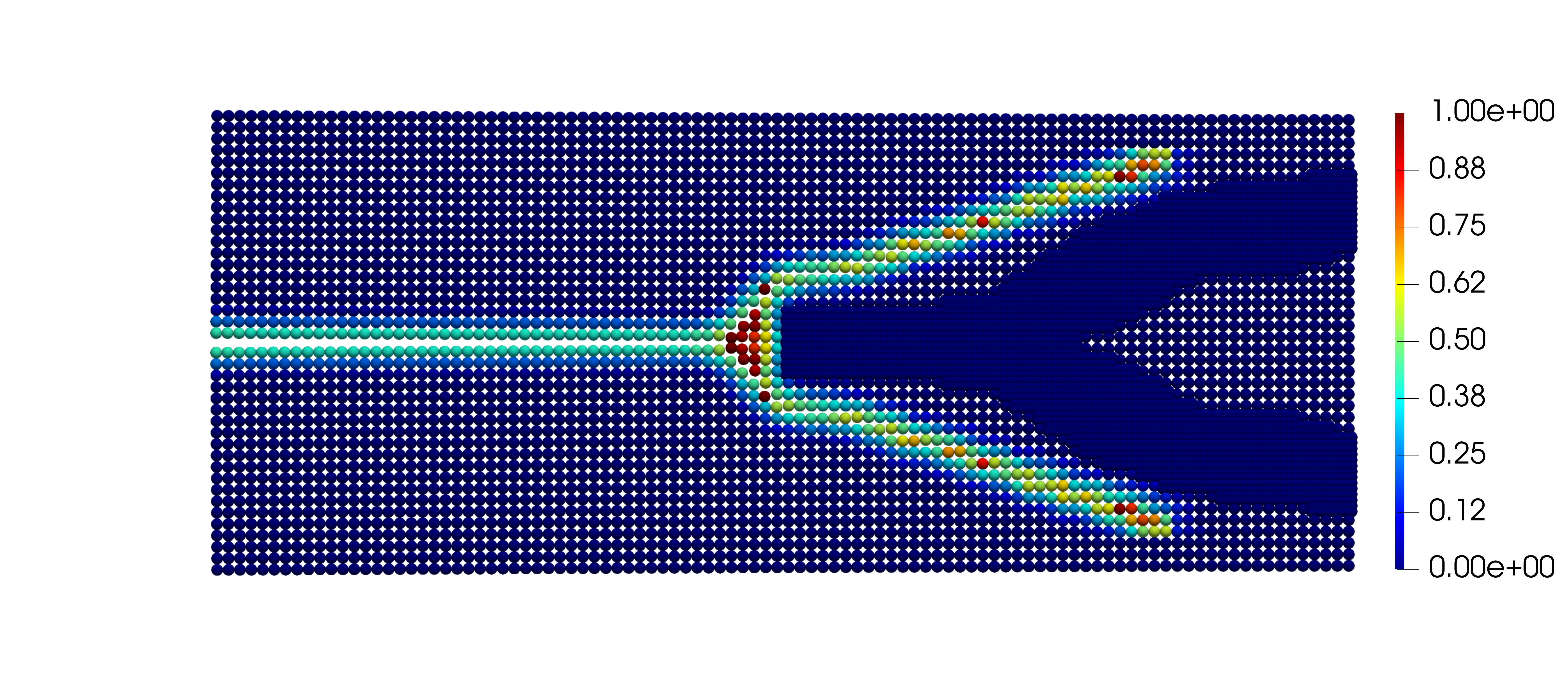}

    \caption{Fracture pattern in a pre-cracked plate at $t = 50\,\mathrm{\mu s}$ obtained using VV scheme with SHBB-PD. The damage field reveals a deviation from the expected crack path, caused by spurious wave reflections at the interface between coarse and refined regions.}

    \label{fig:fracture_pattern_bbpd_50mus}
\end{figure}

In contrast, when the same problem is simulated using the SHBB-PD formulation, a markedly different fracture pattern is obtained, as shown in \cref{fig:fracture_pattern_bbpd_50mus}. In this case, spurious wave reflections arise at the interface between the coarse and refined regions due to the loss of interaction symmetry associated with non-uniform horizons. As a consequence, stress waves originating from the loaded boundaries are partially reflected within the domain, leading to an unphysical redistribution of stresses. This results in a deviation of the crack path from its expected trajectory, with the crack effectively circumventing the refined region as if an artificial barrier were present.

\begin{figure}[!htbp]
    \centering

    \includegraphics{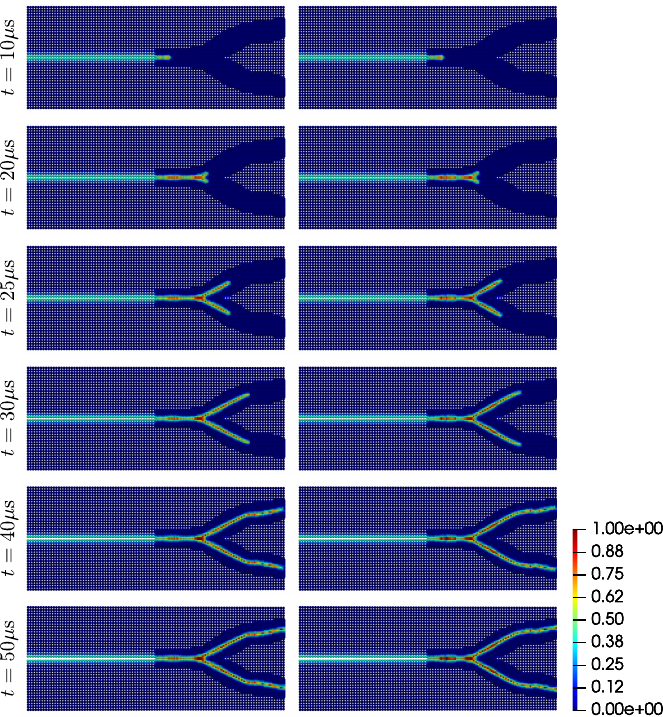}
    \caption{Comparison of fracture patterns in a pre-cracked plate under tension with non-uniform discretisation simulated using (left) the VV and (right) the AVV schemes with DHBB-PD.}
    \label{fig:btt_fracture_patterns_avi_vi_only}
\end{figure}

\cref{fig:btt_fracture_patterns_avi_vi_only} presents a direct comparison between the fracture patterns obtained using the VV and AVV schemes with DHBB-PD. Both approaches predict nearly identical crack trajectories and overall fracture patterns. Although minor differences can be observed in the detailed crack path, these discrepancies are negligible and do not influence the global fracture evolution. Moreover, the non-uniform discretisation, with refinement localized along the expected crack path, achieves a level of accuracy comparable to the uniformly refined case, indicating that a globally fine discretisation is not necessary to resolve the fracture process. This confirms that the asynchronous formulation preserves the essential characteristics of the dynamic fracture process while enabling the use of non-uniform time stepping without introducing spurious numerical artefacts.

In addition to the crack morphology, a quantitative comparison is performed by evaluating the crack tip velocity obtained from the different numerical schemes and discretisations. The crack tip is identified as the point with the maximum horizontal extent in either the upper or lower half of the domain for which the damage exceeds a threshold value of $0.35$, following \cite{haStudiesDynamicCrack2010}. The crack tip position is recorded at discrete time intervals of $f = 2 \mathrm{\mu s}$, and the corresponding velocity is computed using a finite difference approximation,

\begin{equation}
    \vv = \dfrac{\vx^m - \vx^{m-1}}{f}
\end{equation}

\begin{figure}[!htbp]
    \centering

    \includegraphics{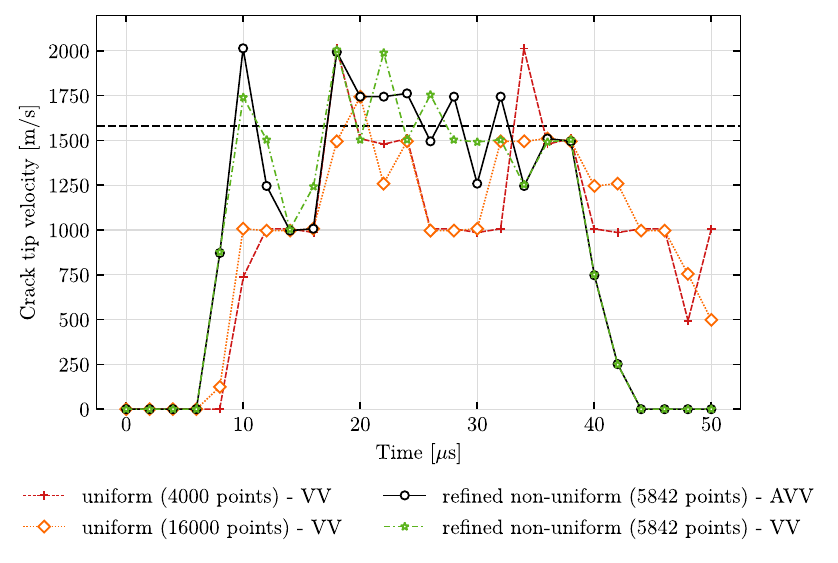}
    \caption{Comparison of crack tip velocities. The dashed horizontal line represents the experimental maximum fracture speed reported by Bowden et al. \cite{bowdenControlledFractureBrittle1967}.}
    \label{fig:crack_velocity_comparison}
\end{figure}

\cref{fig:crack_velocity_comparison} presents the evolution of the crack-tip velocity for the different schemes and discretisation strategies. The results show that crack initiation occurs slightly earlier in the non-uniform discretisations. The onset of crack branching is observed at approximately $21\,\mathrm{\mu s}$ for the uniform discretisation and at around $18\,\mathrm{\mu s}$ for the non-uniform discretisation.

In the non-uniform case, the crack-tip velocity exhibits an initial transient overshoot before converging towards the response obtained with the uniform discretisation. This behaviour is likely caused by internal wave reflections at the interface between the coarse and refined regions, rather than by the asynchronous time integration scheme. Indeed, no additional deviation is observed for the AVV scheme compared with the corresponding VV simulation on the same non-uniform discretisation. Moreover, the crack velocity reaches a second peak shortly before branching, in agreement with the experimental observations of \cite{fieldBrittleFractureIts1971}.

Overall, the predicted crack propagation speeds are in good agreement with experimentally reported maximum fracture velocities, as indicated by the dashed reference line corresponding to the measurements of Bowden et al. \cite{bowdenControlledFractureBrittle1967}. These results demonstrate that the proposed approach captures both the qualitative and quantitative features of dynamic fracture with high fidelity.

The computational efficiency of the AVV integrator relative to the classical VV scheme for the refined non-uniform discretisation is illustrated in \cref{tab:btt_efficiency}. It is important to note that both AVV and VV simulations are performed over the same global time grid, ensuring a consistent basis for comparison. Despite this, the AVV scheme achieves a reduction of approximately $30\%$ in the number of internal force evaluations. Since the evaluation of internal forces constitutes the dominant computational cost in peridynamic simulations, this reduction directly translates into a significant improvement in overall computational efficiency.

\begin{table}[t]
    \centering
    \caption{Comparison of the number of $\Pi_k$ evaluations for the VV and AVV schemes in the pre-cracked plate under tension experiment.}
    \begin{tabular*}{\textwidth}{@{\extracolsep{\fill}} l c}
    \hline
    Method & $\pi^{\mathrm{int}}_k$ evaluations \\ 
    \hline
    VV  & $1.42 \times 10^{7}$ \\
    AVV & $1.01 \times 10^{7}$ \\
    \hline
    \end{tabular*}
    \label{tab:btt_efficiency}
\end{table}

\begin{figure}[!htbp]
    \centering
    \includegraphics{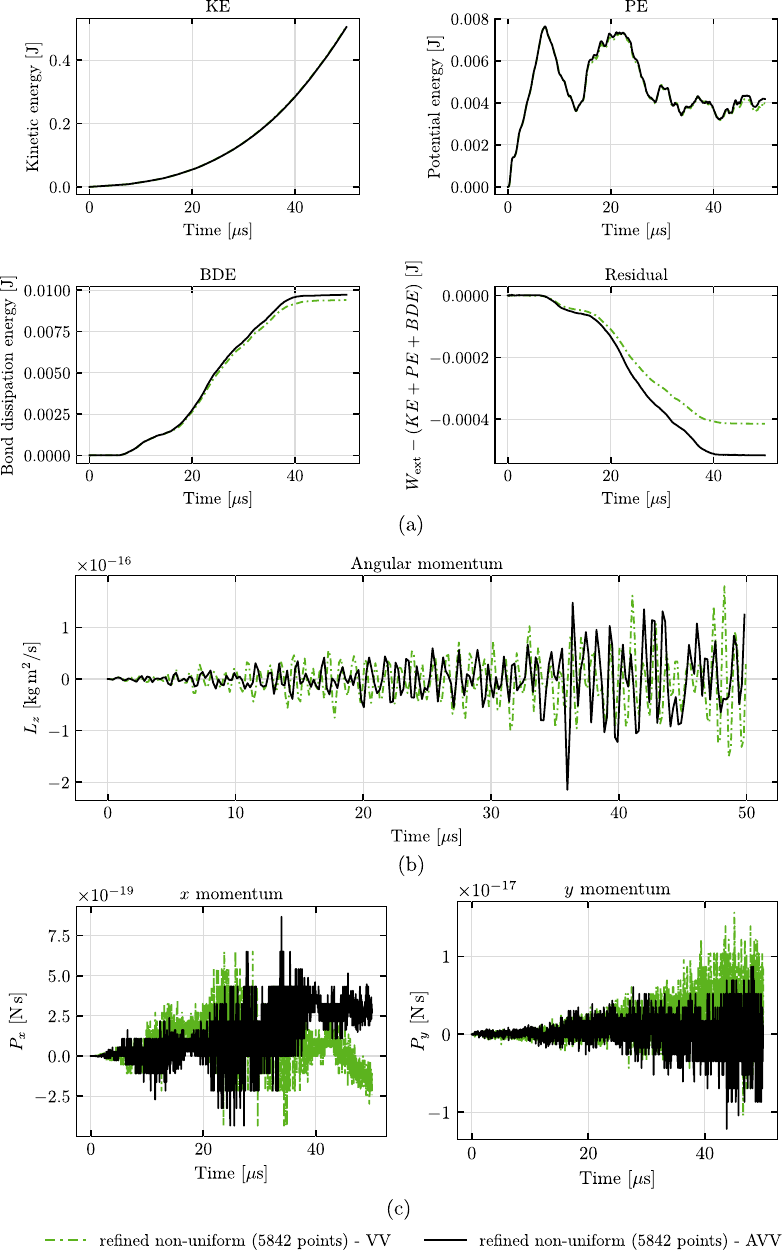}
    \caption{Evolution of the (a) kinetic energy (KE), potential energy (PE), bond-dissipation energy (BDE), and energy-balance residual, defined as $W_{\mathrm{ext}} - (\mathrm{KE}+\mathrm{PE}+\mathrm{BDE})$; (b) angular momentum; and (c) linear momentum along the $x$- and $y$-directions for the pre-cracked plate under tension simulated using the VV and AVV schemes with DHBB-PD.}
    \label{fig:btt_energy_momentum}
\end{figure}

The structure-preserving behaviour of the proposed asynchronous variational integrator is further demonstrated by examining the evolution of the total energy in the system, and the conservation of linear and angular momentum. \cref{fig:btt_energy_momentum} presents the evolution of the total energy, linear momentum, and angular momentum for the pre-cracked plate under tension simulated using both the VV and AVV schemes with DHBB-PD. The results show that both integrators exhibit excellent energy conservation properties, with only minor fluctuations in total energy observed during the simulation. Similarly, linear and angular momentum are well conserved throughout the simulation, with negligible machine precision deviations from their initial values.

\subsection{Kalthoff-Winkler experiment}

\begin{figure}[!htbp]
    \centering

    \includegraphics{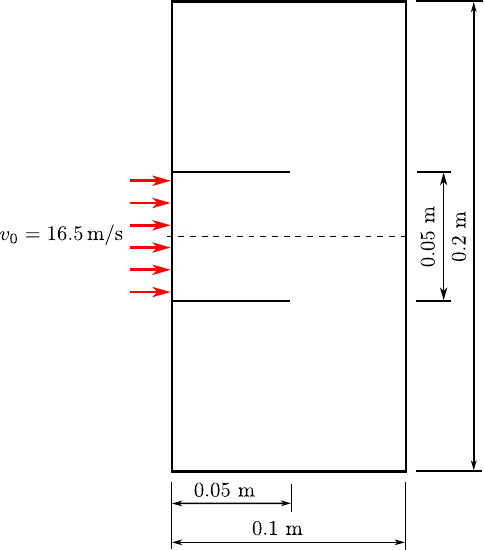}
    \caption{Setup of the Kalthoff-Winkler experiment.}
    \label{fig:kw_diagram}
\end{figure}

To further assess the capability of the proposed framework in capturing dynamic fracture phenomena under impact loading, we consider the well-known Kalthoff--Winkler experiment. This benchmark problem has been widely studied in the fracture mechanics and peridynamics literature due to its ability to reproduce experimentally observed crack initiation angles and branching behaviour under high-rate loading conditions. 

The problem consists of a rectangular plate with two symmetric notches located on one side, as shown in \cref{fig:kw_diagram}. The plate is subjected to impact loading by prescribing an initial velocity on the same face, generating stress waves that propagate through the domain and trigger crack initiation at the notch tips. Under dynamic loading conditions, cracks are known to propagate at an angle of approximately $70^\circ$ with respect to the initial notch direction, as observed in experiments.

The material parameters are chosen as Young's modulus 
$E = 190\,\mathrm{GPa}$, mass density 
$\varrho = 8000\,\mathrm{kg/m^3}$, and critical energy release rate 
$G_c = 22170\,\mathrm{J/m^2}$. The simulations are carried out in two dimensions under plane strain conditions, with Poisson's ratio set to $\nu = 1/4$. To initiate dynamic fracture and reproduce the characteristic crack propagation observed experimentally, an initial velocity of 
$v_0 = 16.5\,\mathrm{m/s}$ is prescribed.

The peridynamic horizon is chosen proportional to the local discretisation size according to $\delta = 3.015\,\Delta xy$
within each region. For temporal stability, a uniform safety factor of $0.5$ is used to determine the time step in all regions. The velocity boundary condition is imposed by prescribing the velocities after completion of the full velocity update step. All simulations are performed up to a final time of 
$90\,\mathrm{\mu s}$. To avoid damage nucleation from the right boundary of the plate, a no-fail layer of width equal to the horizon is introduced along the right edge.

As in the previous example, a non-uniform discretisation is employed, in which the region surrounding the expected crack path is locally refined, while the remaining part of the domain is discretised more coarsely. The coarse region has a spatial resolution of $1\,\mathrm{mm}$, whereas the refined region has a spatial resolution of $0.5\,\mathrm{mm}$ resulting in 28466 total points. The simulations are carried out using both the velocity-Verlet and the asynchronous velocity-Verlet schemes with the DHBB-PD formulation, enabling a direct comparison of the fracture patterns and crack trajectories predicted by the two time integration approaches.

\begin{figure}[!htbp]
    \centering

    \includegraphics{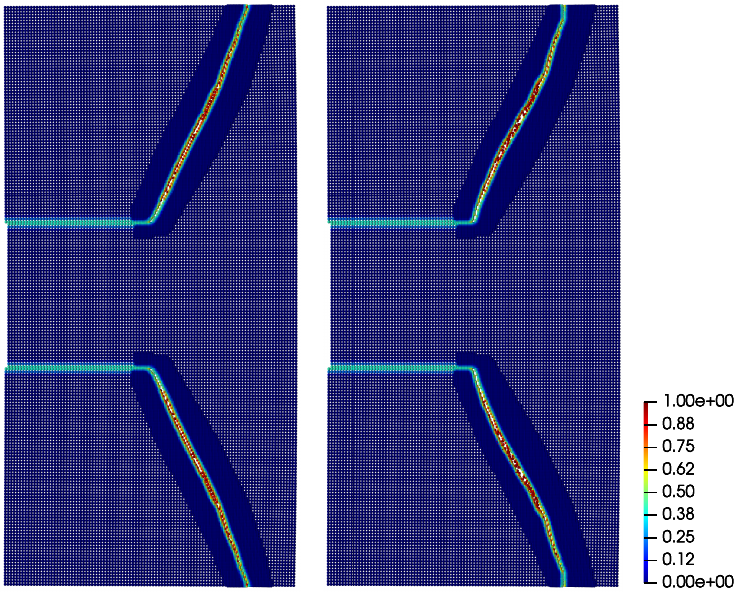}

    \caption{Comparison of fracture patterns in the Kalthoff-Winkler test simulated using (left) the VV and (right) the AVV schemes with DHBB-PD.}
    \label{fig:kw_fracture_patterns_avi_vi_only}
\end{figure}

\begin{figure}[!htbp]4
    \centering

    \includegraphics{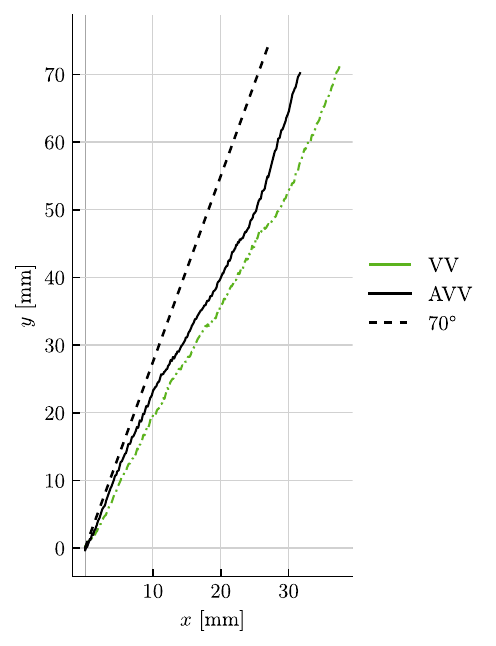}
    \caption{Comparison of the crack tip trajectory obtained using the VV and the AVV schemes with DHBB-PD for the Kalthoff-Winkler experiment. The reference angle of $70^{\degree}$ is indicated by the dashed line.}
    \label{fig:crack_tip_trajectory_comparison}
\end{figure}

\cref{fig:kw_fracture_patterns_avi_vi_only} compares the fracture patterns obtained with the AVV and VV schemes. Both approaches predict nearly identical crack trajectories and overall fracture patterns, with cracks initiating at the notch tips and propagating at an angle close to the characteristic value of $70^{\circ}$. The corresponding crack-tip trajectories are shown in \cref{fig:crack_tip_trajectory_comparison}, where both integrators are compared against the reference propagation angle. The AVV scheme captures the crack trajectory with high fidelity and closely follows the reference angle throughout the simulation. The VV scheme predicts a similar trajectory, although a slightly larger deviation from the reference angle is observed during the later stages of crack propagation.

Overall, these results demonstrate that the proposed asynchronous variational integrator is capable of accurately capturing the complex fracture patterns and crack trajectories observed in the Kalthoff--Winkler experiment, while permitting non-uniform time stepping without introducing spurious numerical artefacts. This leads to a substantial reduction in computational cost. As reported in \cref{tab:kw_efficiency}, the number of internal force evaluations is reduced by approximately $30\%$ when using the AVV scheme compared with the VV scheme. This reduction is particularly relevant for the present problem, where a high spatial resolution is required to accurately resolve the complex fracture process.

\begin{table}[!htbp]
    \centering
    \caption{Comparison of the number of $\Pi_k$ evaluations for the VV and AVV schemes in the Kalthoff-Winkler experiment.}
    \begin{tabular*}{\textwidth}{@{\extracolsep{\fill}} l c}
    \hline
    Method & $\pi^{\mathrm{int}}_k$ evaluations \\ 
    \hline
    VV  & $4.63 \times 10^{7}$ \\
    AVV & $3.23 \times 10^{7}$ \\
    \hline
    \end{tabular*}
    \label{tab:kw_efficiency}
\end{table}


\section{Conclusions} \label{sec:conclusions}

In this work, a unified variational framework for bond-based peridynamics with spatially varying horizons has been developed. Starting from the Lagrange-d'Alembert principle, the governing equations of motion were derived in a consistent variational setting, revealing that asymmetric interactions arising from non-uniform horizons naturally give rise to the dual-horizon formulation. This provides a rigorous mechanical foundation for dual-horizon bond-based peridynamics, which was previously introduced as a corrective construction.

Building on this variational formulation, asynchronous variational integrators were constructed to enable the use of spatially varying time step sizes. The resulting integration schemes preserve key geometric and physical properties of the system while allowing for efficient time integration in problems involving localized features such as cracks.

The numerical examples demonstrate the effectiveness of the proposed approach. In the wave propagation problem, the SHBB-PD formulation with non-uniform discretisations together with spatially varying horizons exhibits spurious reflections due to asymmetric interactions, whereas the DHBB-PD formulation eliminates these artefacts and restores physically consistent wave transmission. In dynamic fracture simulations, including the pre-cracked plate and the Kalthoff–Winkler experiment, the proposed framework accurately captures crack initiation, propagation, and branching behaviour. The results show that non-uniform discretisations, when combined with the dual-horizon formulation, do not introduce spurious numerical effects and can achieve accuracy comparable to uniformly refined discretisations.

Furthermore, the asynchronous variational integrator demonstrates significant computational advantages. By allowing smaller time steps only in regions requiring higher resolution, the AVV scheme reduces the number of internal force evaluations by approximately $30\%$ compared to the classical VV scheme, while maintaining accuracy and stability. This highlights the potential of the proposed approach for large-scale dynamic fracture simulations.

Overall, the present work establishes a consistent variational foundation for bond-based peridynamics with variable horizons and introduces a structure-preserving asynchronous time integration strategy. The framework is general and can be extended to more advanced peridynamic models, including ordinary and non-ordinary state-based formulations. Future work will focus on extending the approach to and exploring fully adaptive spatio-temporal refinement strategies, and also the parallel implementation of the AVV scheme for large-scale simulations.

\section*{CRediT authorship contribution statement}

\textbf{Prateek Prateek:} Conceptualization, Methodology, Software, Validation, Formal analysis, Investigation, Writing - original draft, Visualization. \textbf{Giuseppe Capobianco:} Conceptualization, Methodology, Formal analysis, Writing - review \& editing, Supervision. \textbf{Kai Partmann:} Methodology, Validation, Writing - review \& editing. \textbf{Kerstin Weinberg:} Methodology, Validation, Writing - review \& editing. \textbf{Michael Ortiz:} Methodology, Validation, Writing - review \& editing. \textbf{Sigrid Leyendecker:} Conceptualization, Methodology, Writing - review \& editing, Supervision, Project administration, Funding acquisition.

\section*{Declaration of competing interests}

The authors declare that they have no conflict of interest.

\section*{Data Availability}

The datasets generated during and/or analysed during the current study are available from the corresponding author on reasonable request.

\section*{Acknowledgements}

This research was funded by the Deutsche Forschungsgemeinschaft (DFG, German Research Foundation) - 377472739/GRK 2423/2-2023. The authors are very grateful for this support.

\section*{Declaration of generative AI and AI-assisted technologies in the manuscript preparation process}
During the preparation of this work, the authors used ChatGPT to paraphrase drafted sentences, provide wording suggestions, and assist with language, punctuation, and grammar. After using this tool, the authors reviewed and edited the content as needed and take full responsibility for the content of the published article.

\newpage

\printbibliography

@article{bowdenControlledFractureBrittle1967,
  title = {Controlled {{Fracture}} of {{Brittle Solids}} and {{Interruption}} of {{Electrical Current}}},
  author = {Bowden, F. P. and Brunton, J. H. and Field, J. E. and Heyes, A. D.},
  year = 1967,
  month = oct,
  journal = {Nature},
  volume = {216},
  number = {5110},
  pages = {38--42},
  issn = {1476-4687},
  doi = {10.1038/216038a0},
  urldate = {2026-04-19},
  copyright = {1967 Springer Nature Limited},
  langid = {english}
}

@article{chenPeridynamicBondassociatedCorrespondence2019,
  title = {Peridynamic Bond-Associated Correspondence Model: {{Stability}} and Convergence Properties},
  shorttitle = {Peridynamic Bond-Associated Correspondence Model},
  author = {Chen, Hailong and Spencer, Benjamin W.},
  year = 2019,
  journal = {Int. J. Numer. Methods Eng.},
  volume = {117},
  number = {6},
  pages = {713--727},
  issn = {1097-0207},
  doi = {10.1002/nme.5973},
  urldate = {2026-03-22},
  langid = {english}
}

@article{dipasqualeCrackPropagationAdaptive2014,
  title = {Crack Propagation with Adaptive Grid Refinement in {{2D}} Peridynamics},
  author = {Dipasquale, Daniele and Zaccariotto, Mirco and Galvanetto, Ugo},
  year = 2014,
  month = nov,
  journal = {Int. J. Fract.},
  volume = {190},
  number = {1},
  pages = {1--22},
  issn = {1573-2673},
  doi = {10.1007/s10704-014-9970-4},
  urldate = {2026-02-23},
  langid = {english}
}

@article{fieldBrittleFractureIts1971,
  title = {Brittle Fracture: {{Its}} Study and Application},
  shorttitle = {Brittle Fracture},
  author = {Field, J. E.},
  year = 1971,
  month = jan,
  journal = {Contemp. Phys.},
  volume = {12},
  number = {1},
  pages = {1--31},
  issn = {0010-7514, 1366-5812},
  doi = {10.1080/00107517108205103},
  urldate = {2026-04-19},
  langid = {english}
}

@article{gailVariationalMultirateIntegration2016,
  title = {Variational Multirate Integration in Multi-Body Dynamics},
  author = {Gail, Tobias and Leyendecker, Sigrid and {Ober-Bl{\"o}baum}, Sina},
  year = 2016,
  journal = {Proc. Appl. Math. Mech.},
  volume = {16},
  number = {1},
  pages = {53--54},
  issn = {1617-7061},
  doi = {10.1002/pamm.201610015},
  urldate = {2026-06-26},
  langid = {english}
}

@article{gerstlePeridynamicModelingConcrete2007,
  title = {Peridynamic Modeling of Concrete Structures},
  author = {Gerstle, Walter and Sau, Nicolas and Silling, Stewart},
  year = 2007,
  month = jul,
  journal = {Nucl. Eng. Des.},
  series = {18th {{International Conference}} on {{Structural Mechanics}} in {{Nuclear Engineering}}},
  volume = {237},
  number = {12},
  pages = {1250--1258},
  issn = {0029-5493},
  doi = {10.1016/j.nucengdes.2006.10.002},
  urldate = {2026-03-22}
}

@article{haStudiesDynamicCrack2010,
  title = {Studies of Dynamic Crack Propagation and Crack Branching with Peridynamics},
  author = {Ha, Youn Doh and Bobaru, Florin},
  year = 2010,
  month = mar,
  journal = {Int. J. Fract.},
  volume = {162},
  number = {1},
  pages = {229--244},
  issn = {1573-2673},
  doi = {10.1007/s10704-010-9442-4},
  urldate = {2025-10-07},
  langid = {english}
}

@article{jadhavNewApproachAsynchronous2025,
  title = {A {{New Approach}} to {{Asynchronous Variational Integrators}} for a {{Phase Field Model}} of {{Dynamic Fracture}}},
  author = {Jadhav, Deepak B. and Phansalkar, Dhananjay and Weinberg, Kerstin and Ortiz, Michael and Leyendecker, Sigrid},
  year = 2025,
  journal = {Int. J. Numer. Methods Eng.},
  volume = {126},
  number = {6},
  pages = {e70025},
  issn = {1097-0207},
  doi = {10.1002/nme.70025},
  urldate = {2025-06-26},
  langid = {english}
}

@article{jadhavSpatiotemporallyAdaptiveAsynchronous2026,
  title = {A Spatio-Temporally Adaptive Asynchronous Variational Integrator for a Phase Field Model of Dynamic Fracture},
  author = {Jadhav, Deepak B. and Phansalkar, Dhananjay and Weinberg, Kerstin and Ortiz, Michael and Leyendecker, Sigrid},
  year = 2026,
  month = apr,
  journal = {Mech. Mater.},
  volume = {215},
  pages = {105604},
  issn = {0167-6636},
  doi = {10.1016/j.mechmat.2026.105604},
  urldate = {2026-03-23}
}

@article{javaheriHigherOrderApproximationsStabilizing2022,
  title = {Higher-{{Order Approximations}} for {{Stabilizing Zero-Energy Modes}} in {{Non-Ordinary State-Based Peridynamics Models}}},
  author = {Javaheri, Iman and Luo, Jiangyi and Lakshmanan, Aaditya and Sundararaghavan, Veera},
  year = 2022,
  month = aug,
  journal = {AIAA J.},
  volume = {60},
  number = {8},
  pages = {4906--4922},
  publisher = {{American Institute of Aeronautics and Astronautics}},
  issn = {0001-1452},
  doi = {10.2514/1.J061453},
  urldate = {2026-03-22}
}

@incollection{leitzVariationalLieGroup2014,
  title = {Variational {{Lie Group Formulation}} of {{Geometrically Exact Beam Dynamics}}: {{Synchronous}} and {{Asynchronous Integration}}},
  shorttitle = {Variational {{Lie Group Formulation}} of {{Geometrically Exact Beam Dynamics}}},
  booktitle = {Multibody {{Dynamics}}: {{Computational Methods}} and {{Applications}}},
  author = {Leitz, Thomas and {Ober-Bl{\"o}baum}, Sina and Leyendecker, Sigrid},
  editor = {Terze, Zdravko},
  year = 2014,
  pages = {175--203},
  publisher = {Springer International Publishing},
  address = {Cham},
  doi = {10.1007/978-3-319-07260-9_8},
  urldate = {2026-03-22},
  isbn = {978-3-319-07260-9},
  langid = {english}
}

@article{leSurfaceCorrectionsPeridynamic2018,
  title = {Surface Corrections for Peridynamic Models in Elasticity and Fracture},
  author = {Le, Q. V. and Bobaru, F.},
  year = 2018,
  month = apr,
  journal = {Comput. Mech.},
  volume = {61},
  number = {4},
  pages = {499--518},
  issn = {1432-0924},
  doi = {10.1007/s00466-017-1469-1},
  urldate = {2026-03-22},
  langid = {english}
}

@article{lewAsynchronousVariationalIntegrators2003,
  title = {Asynchronous {{Variational Integrators}}},
  author = {Lew, A. and Marsden, J. E. and Ortiz, M. and West, M.},
  year = 2003,
  month = apr,
  journal = {Arch. Ration. Mech. Anal.},
  volume = {167},
  number = {2},
  pages = {85--146},
  issn = {1432-0673},
  doi = {10.1007/s00205-002-0212-y},
  urldate = {2026-02-20},
  langid = {english}
}

@phdthesis{lewThesisCaltech2003,
  title = {Variational Time Integrators in Computational Solid Mechanics},
  author = {Lew, Andrew},
  year = 2003,
  address = {Pasadena, California},
  doi = {10.7907/6C74-GC16},
  school = {California Institute of Technology}
}

@incollection{leyendeckerVariationalApproachMultirate2013,
  title = {A {{Variational Approach}} to {{Multirate Integration}} for {{Constrained Systems}}},
  booktitle = {Multibody {{Dynamics}}: {{Computational Methods}} and {{Applications}}},
  author = {Leyendecker, Sigrid and {Ober-Bl{\"o}baum}, Sina},
  editor = {Samin, Jean-Claude and Fisette, Paul},
  year = 2013,
  pages = {97--121},
  publisher = {Springer Netherlands},
  address = {Dordrecht},
  doi = {10.1007/978-94-007-5404-1_5},
  urldate = {2026-06-25},
  isbn = {978-94-007-5404-1},
  langid = {english}
}

@misc{lishkovaMultirateVariationalApproach2020,
  title = {A Multirate Variational Approach to Simulation and Optimal Control for Flexible Spacecraft},
  author = {Lishkova, Yana and {Ober-Bl{\"o}baum}, Sina and Cannon, Mark and Leyendecker, Sigrid},
  year = 2020,
  month = oct,
  number = {arXiv:2009.05873},
  eprint = {2009.05873},
  primaryclass = {math.OC},
  publisher = {arXiv},
  doi = {10.48550/arXiv.2009.05873},
  urldate = {2026-06-26},
  archiveprefix = {arXiv}
}

@techreport{littlewoodEstimationCriticalTime2014,
  title = {Estimation of the {{Critical Time Step}} for {{Peridynamic Models}}.},
  author = {Littlewood, David John and Thomas, Jesse David and Shelton, Timothy},
  year = 2014,
  month = jun,
  number = {SAND2014-15123PE},
  institution = {Sandia National Laboratories (SNL-NM), Albuquerque, NM (United States)},
  urldate = {2025-10-28},
  langid = {english}
}

@article{liuOrdinaryStatebasedPeridynamic2018,
  title = {An Ordinary State-Based Peridynamic Model for the Fracture of Zigzag Graphene Sheets},
  author = {Liu, Xuefeng and He, Xiaoqiao and Wang, Jinbao and Sun, Ligang and Oterkus, Erkan},
  year = 2018,
  month = sep,
  journal = {Proc. R. Soc. Math. Phys. Eng. Sci.},
  volume = {474},
  number = {2217},
  pages = {20180019},
  issn = {1364-5021},
  doi = {10.1098/rspa.2018.0019},
  urldate = {2026-03-22}
}

@book{madenciPeridynamicTheoryIts2014,
  title = {Peridynamic {{Theory}} and {{Its Applications}}},
  author = {Madenci, Erdogan and Oterkus, Erkan},
  year = 2014,
  publisher = {Springer},
  address = {New York, NY},
  doi = {10.1007/978-1-4614-8465-3},
  urldate = {2026-03-22},
  copyright = {https://www.springernature.com/gp/researchers/text-and-data-mining},
  isbn = {978-1-4614-8464-6 978-1-4614-8465-3},
  langid = {english}
}

@article{MarsdenWest2001,
  title = {Discrete {{Mechanics}} and {{Variational Integrators}}},
  author = {Marsden, Jerrold E. and West, Matthew},
  year = 2001,
  volume = {10},
  pages = {357--514},
  doi = {10.1017/S096249290100006X}
}

@techreport{mitchellNonlocalOrdinaryStatebased2011,
  title = {A Nonlocal, Ordinary, State-Based Plasticity Model for Peridynamics.},
  author = {Mitchell, John},
  year = 2011,
  month = may,
  number = {SAND2011-3166, 1018475},
  pages = {SAND2011-3166, 1018475},
  doi = {10.2172/1018475},
  urldate = {2026-03-22},
  langid = {english}
}

@article{niuAsynchronousVariationalIntegrator2024,
  title = {An {{Asynchronous Variational Integrator}} for {{Contact Problems Involving Elastoplastic Solids}}},
  author = {Niu, Zongwu and Wang, Zixiao and Shen, Yongxing},
  year = 2024,
  month = apr,
  journal = {Acta Mech. Solida Sin.},
  volume = {37},
  number = {2},
  pages = {305--315},
  issn = {1860-2134},
  doi = {10.1007/s10338-023-00456-2},
  urldate = {2026-03-22},
  langid = {english}
}

@misc{ober-blobaumVariationalMultirateIntegrators2024,
  title = {Variational Multirate Integrators},
  author = {{Ober-Bl{\"o}baum}, Sina and Wenger, Theresa and Gail, Tobias and Leyendecker, Sigrid},
  year = 2024,
  month = jun,
  number = {arXiv:2406.12991},
  eprint = {2406.12991},
  primaryclass = {math.NA},
  publisher = {arXiv},
  doi = {10.48550/arXiv.2406.12991},
  urldate = {2026-06-25},
  archiveprefix = {arXiv}
}

@article{partmannPeridynamicComputationsWave2024,
  title = {Peridynamic Computations of Wave Propagation and Reflection at Material Interfaces},
  author = {Partmann, Kai and Dienst, Manuel and Weinberg, Kerstin},
  year = 2024,
  month = sep,
  journal = {Arch. Appl. Mech.},
  volume = {94},
  number = {9},
  pages = {2405--2416},
  issn = {1432-0681},
  doi = {10.1007/s00419-024-02646-x},
  urldate = {2025-11-13},
  langid = {english}
}

@article{rakiciDiscreteSurfaceCorrection2023,
  title = {A Discrete Surface Correction Method for Bond-Based Peridynamics},
  author = {Rakici, Semsi and Kim, Jinseok},
  year = 2023,
  month = jun,
  journal = {Eng. Anal. Bound. Elem.},
  volume = {151},
  pages = {115--135},
  issn = {09557997},
  doi = {10.1016/j.enganabound.2023.02.041},
  urldate = {2025-06-24},
  langid = {english}
}

@article{renDualhorizonPeridynamics2016,
  title = {Dual-Horizon Peridynamics},
  author = {Ren, Huilong and Zhuang, Xiaoying and Cai, Yongchang and Rabczuk, Timon},
  year = 2016,
  journal = {Int. J. Numer. Methods Eng.},
  volume = {108},
  number = {12},
  pages = {1451--1476},
  issn = {1097-0207},
  doi = {10.1002/nme.5257},
  urldate = {2025-11-04},
  copyright = {Copyright \copyright{} 2016 John Wiley \& Sons, Ltd.},
  langid = {english}
}

@article{renDualhorizonPeridynamicsStable2017,
  title = {Dual-Horizon Peridynamics: {{A}} Stable Solution to Varying Horizons},
  shorttitle = {Dual-Horizon Peridynamics},
  author = {Ren, Huilong and Zhuang, Xiaoying and Rabczuk, Timon},
  year = 2017,
  month = may,
  journal = {Comput. Methods Appl. Mech. Eng.},
  volume = {318},
  pages = {762--782},
  issn = {0045-7825},
  doi = {10.1016/j.cma.2016.12.031},
  urldate = {2025-11-12}
}

@article{sillingMeshfreeMethodBased2005,
  title = {A Meshfree Method Based on the Peridynamic Model of Solid Mechanics},
  author = {Silling, S.A. and Askari, E.},
  year = 2005,
  month = jun,
  journal = {Comput. Struct.},
  volume = {83},
  number = {17-18},
  pages = {1526--1535},
  issn = {00457949},
  doi = {10.1016/j.compstruc.2004.11.026},
  urldate = {2025-06-24},
  copyright = {https://www.elsevier.com/tdm/userlicense/1.0/},
  langid = {english}
}

@article{sillingPeridynamicStatesConstitutive2007,
  title = {Peridynamic {{States}} and {{Constitutive Modeling}}},
  author = {Silling, S. A. and Epton, M. and Weckner, O. and Xu, J. and Askari, E.},
  year = 2007,
  month = aug,
  journal = {J. Elast.},
  volume = {88},
  number = {2},
  pages = {151--184},
  issn = {1573-2681},
  doi = {10.1007/s10659-007-9125-1},
  urldate = {2025-10-23},
  langid = {english}
}

@article{sillingReformulationElasticityTheory2000,
  title = {Reformulation of Elasticity Theory for Discontinuities and Long-Range Forces},
  author = {Silling, S. A.},
  year = 2000,
  month = jan,
  journal = {J. Mech. Phys. Solids},
  volume = {48},
  number = {1},
  pages = {175--209},
  issn = {0022-5096},
  doi = {10.1016/S0022-5096(99)00029-0},
  urldate = {2025-10-23}
}

@article{sillingStabilityPeridynamicCorrespondence2017,
  title = {Stability of Peridynamic Correspondence Material Models and Their Particle Discretizations},
  author = {Silling, S. A.},
  year = 2017,
  month = aug,
  journal = {Comput. Methods Appl. Mech. Eng.},
  volume = {322},
  pages = {42--57},
  issn = {0045-7825},
  doi = {10.1016/j.cma.2017.03.043},
  urldate = {2026-03-22}
}

@article{sillingVariableHorizonPeridynamic2015,
  title = {Variable Horizon in a Peridynamic Medium},
  author = {Silling, Stewart and Littlewood, David and Seleson, Pablo},
  year = 2015,
  month = dec,
  journal = {J. Mech. Mater. Struct.},
  volume = {10},
  number = {5},
  pages = {591--612},
  issn = {1559-3959, 1559-3959},
  doi = {10.2140/jomms.2015.10.591},
  urldate = {2026-02-12},
  langid = {english}
}

@article{trageserBondBasedPeridynamicsTale2020,
  title = {Bond-{{Based Peridynamics}}: A {{Tale}} of {{Two Poisson}}'s {{Ratios}}},
  shorttitle = {Bond-{{Based Peridynamics}}},
  author = {Trageser, Jeremy and Seleson, Pablo},
  year = 2020,
  month = sep,
  journal = {J. Peridynamics Nonlocal Model.},
  volume = {2},
  number = {3},
  pages = {278--288},
  issn = {2522-8978},
  doi = {10.1007/s42102-019-00021-x},
  urldate = {2025-10-28},
  langid = {english}
}

@article{vougaAsynchronousVariationalContact2011,
  title = {Asynchronous Variational Contact Mechanics},
  author = {Vouga, E. and Harmon, D. and Tamstorf, R. and Grinspun, E.},
  year = 2011,
  month = jun,
  journal = {Comput. Methods Appl. Mech. Eng.},
  volume = {200},
  number = {25},
  pages = {2181--2194},
  issn = {0045-7825},
  doi = {10.1016/j.cma.2011.03.010},
  urldate = {2026-03-22}
}

@article{warrenNonordinaryStatebasedPeridynamic2009,
  title = {A Non-Ordinary State-Based Peridynamic Method to Model Solid Material Deformation and Fracture},
  author = {Warren, Thomas L. and Silling, Stewart A. and Askari, Abe and Weckner, Olaf and Epton, Michael A. and Xu, Jifeng},
  year = 2009,
  month = mar,
  journal = {Int. J. Solids Struct.},
  volume = {46},
  number = {5},
  pages = {1186--1195},
  issn = {0020-7683},
  doi = {10.1016/j.ijsolstr.2008.10.029},
  urldate = {2026-03-22}
}

@article{wolffAsynchronousVariationalIntegration2013,
  title = {Asynchronous Variational Integration Using Continuous Assumed Gradient Elements},
  author = {Wolff, Sebastian and Bucher, Christian},
  year = 2013,
  month = mar,
  journal = {Comput. Methods Appl. Mech. Eng.},
  volume = {255},
  pages = {158--166},
  issn = {0045-7825},
  doi = {10.1016/j.cma.2012.11.004},
  urldate = {2026-03-22}
}

@phdthesis{wolffThesisAVI,
  title = {Asychronous Variational Integration of Structural Collision Dynamics},
  author = {Wolff, Sebastian},
  school = {TU Wien}
}

\end{document}